\documentclass{article}
\usepackage{geometry}
\usepackage{amsmath}
\usepackage{amssymb}
\usepackage{amsthm}
\usepackage{graphicx}
\usepackage{booktabs}
\usepackage[numbers]{natbib}
\usepackage{setspace}
\usepackage{color}
\usepackage{caption}   
\usepackage{comment}

\newtheorem{theorem}{Theorem}

\title{\Large Estimand-based Inference in Presence of Long-Term Survivors}

\author{
    Yi-Cheng Tai\textsuperscript{1,2},
    Weijing Wang\textsuperscript{2}\thanks{Corresponding Author: Email: wjwang@stat.nycu.edu.tw},
    Martin T. Wells\textsuperscript{1}
}
\date{}

\begin{document}
\maketitle

\vspace{-10mm} 

\begin{flushleft}
\textsuperscript{1} Department of Statistics and Data Science, Cornell University, Ithaca, NY, USA\\
\textsuperscript{2} Institute of Statistics, National Yang Ming Chiao Tung University, Taiwan, ROC
\end{flushleft}

\begin{abstract}
In this article, we develop nonparametric inference methods for comparing survival data across two samples, which are beneficial for clinical trials of novel cancer therapies where long-term survival is a critical outcome. These therapies, including immunotherapies or other advanced treatments, aim to establish durable effects. They often exhibit distinct survival patterns such as crossing or delayed separation and potentially leveling-off at the tails of survival curves, clearly violating the proportional hazards assumption and rendering the hazard ratio inappropriate for measuring treatment effects. The proposed methodology utilizes the mixture cure framework to separately analyze the cure rates of long-term survivors and the survival functions of susceptible individuals. We evaluate a nonparametric estimator for the susceptible survival function in the one-sample setting. Under sufficient follow-up, it is expressed as a location-scale-shift variant of the Kaplan-Meier (KM) estimator. It retains several desirable features of the KM estimator, including inverse-probability-censoring weighting, product-limit estimation, self-consistency, and nonparametric efficiency. In scenarios of insufficient follow-up, it can easily be adapted by incorporating a suitable cure rate estimator.
In the two-sample setting, besides using the difference in cure rates to measure the long-term effect, we propose a graphical estimand to compare the relative treatment effects on susceptible subgroups. This process, inspired by Kendall's tau, compares the order of survival times among susceptible individuals. The proposed methods' large-sample properties are derived for further inference, and the finite-sample properties are examined through extensive simulation studies. The proposed methodology is applied to analyze the digitized data from the CheckMate 067 immunotherapy clinical trial.
\end{abstract}

\noindent \textbf{Keywords:} Cure mixture model, Estimand, Immunotherapy, Insufficient follow-up, Kaplan-Meier estimator, Kendall's tau, Nonproportional hazards, Self-consistency, Sufficient follow-up, Two-sample comparison

\maketitle

\section{Introduction}
\label{sec1}
In clinical trials for novel cancer therapies, these treatments often exhibit distinct effect profiles compared to conventional treatments such as chemotherapy, likely due to their different pharmacological mechanisms \cite{chen2013statistical, tsimberidou2015targeted,Schirrmacher2019}. For instance, immunotherapy stimulates the immune system to attack cancer cells, while targeted therapies disrupt cancer cell growth by focusing on specific genetic alterations or proteins, thus minimizing harm to healthy cells. Chemotherapy, often used as the control group in these trials, primarily aims to directly destroy rapidly dividing cancer cells. These varying mechanisms of action lead to different temporal effects on patient outcomes, which are evident in the survival curves. The crossing or delayed separation and subsequent leveling-off in the tails of survival curves, as shown in many reports, clearly violate the proportional hazards assumption.
In such scenarios, the log-rank test may lose power, and the hazard ratio becomes an inappropriate measure for evaluating treatment effects \cite{uno2014moving, stensrud2019limitations, bartlett2020hazards,Bardo2024}.
While novel therapies aim to improve long-term survival, the presence of both long-term survivors and patients who do not respond favorably to treatments introduces complexity into clinical trial analyses \cite{mcdermott2014durable, quinn2020current}. The heterogeneity in patient responses often makes interpreting trial results challenging \cite{bartlett2020hazards, stensrud2018limitations}.
In accordance with the ICH guidelines \cite{ICH2019E9R1}, formulating suitable estimands for clinical trials involving long-term survivors is crucial. This is essential for the accurate analysis and interpretation of trial results, particularly when traditional statistical methods are inadequate.
 
Empirical evidence indicates that the long-term benefits of promising novel therapies might be limited to a subset of patients \cite{quinn2020current}. This observation aligns with the mixture formulation approach commonly used in analyzing survival data with a cure fraction \cite{felizzi2021mixture}.  
It assumes that there are two groups, one with the potential to be ``cured,''  and the other ``uncured'' or susceptible subjects. In this context, the term 'cure' does not necessarily imply the complete eradication of the disease but may instead represent a long-term survival probability that indicates no further risk of the event of interest. 
Therefore, we adopt this framework to develop suitable estimands for separately measuring the treatment effects in the two sub-populations.

Nonparametric inference of the mixture cure model requires that the follow-up period is sufficiently long to ensure the potential observation of maximum lifetimes in the susceptible sub-population, necessary for the purpose of identifiability \cite{maller1996survival}. In the medical field, landmark trials that aim to establish the long-term efficacy of treatments often include results from extended follow-up periods.
An example is the CheckMate 067 trial, which enrolled 1,296 patients across 137 sites globally and extended its follow-up to at least 6.5 years, providing crucial insights into the long-term outcomes and safety of the treatments \cite{hodi2018nivolumab, Wolchok2022}. On the other hand, statistical methods for scenarios of insufficient follow-up have also been developed. Escobar-Bach and colleagues have utilized extrapolation techniques from extreme value theory to reduce the bias due to missing tail information \cite{EscobarBach2019, EscobarBach2023}. Furthermore, the paper by Maller et al. reviews recent progress in cure mixture models \cite{Maller2024}.

In the presence of covariates, various approaches to analyzing the cure mixture model have been explored. \textit{Statistical Methods in Medical Research} featured a special issue on this topic, edited by Balakrishnan in 2017 \cite{Balakrishnan2017}.
In a two-sample setting, we assess treatment effects by examining differences across key measures. The difference in cure fractions highlights the impact on long-term survivors, while comparisons of survival times among susceptible sub-groups allow us to evaluate the treatment's efficacy for those who do not achieve cure status.
For the latter comparison, the Cox-TEL method, which utilizes a Taylor expansion technique to bridge Cox proportional hazards (PH) and PH cure models for data with long-term survival, adjusts the hazard ratio to more accurately quantify the treatment effect among the two susceptible groups \cite{lin2020changing, hsu2021development}. However, the Cox-TEL approach relies on the proportional hazards assumption for the two susceptible groups, an assumption that may be applicable only in certain limited clinical trial scenarios.


In Section~\ref{sec2}, we first review the cure mixture framework and related existing results. We then focus on a location-scale-shift variant of the Kaplan-Meier (KM) estimator to estimate the susceptible survival function. With sufficient follow-up, we demonstrate that this estimator possesses several useful representations that parallel those of the KM estimator. These include the product-limit form, the Inverse Probability of Censoring Weighting (IPCW) expression, and a self-consistency equation. We also establish the theoretical properties of this latency survival estimator, including weak convergence and nonparametric efficiency. If follow-up is not sufficient, we demonstrate that the susceptible survival function can be estimated by incorporating the cure rate estimator proposed by Escobar-Bach and Van Keilegom \cite{EscobarBach2019}. Their method involves adding a compensating component to the tail of the KM estimator.

Section 3 examines the cure mixture formulation within a two-sample framework. Beyond assessing long-term survivor outcomes by comparing differences in cure rates between treatment groups, we introduce a graphical estimand that captures the temporal impact of treatments on susceptible groups. The proposed method is a modification of the approach originally developed by Tai et al. \cite{tai2023twosample}. This plot offers a clear interpretation of the relative treatment effect on the susceptible groups over time, which is preferred to a single number summary such as an average hazard rate or restricted mean survival rate \cite{Bardo2024}. Our approach does not make any additional assumptions about the relationship between the groups under comparison, enhancing its versatility compared to the Cox-TEL approach.
An extensive simulation study examining the finite-sample properties of the proposed methodology is presented in Section 4, and an application of this methodology to the CheckMate 067 clinical trial is detailed in Section 5. The technical details, proofs of theorems, and additional numerical results are provided in the Supplementary Material.

\section{Estimation of Susceptible Survival Function}\label{sec2}
In the one-sample setting, let $T$ be the failure time with the survival function $S(t) = \mbox{Pr}(T > t)$. Define $\xi$ as the  indicator of susceptibility such that $T < \infty$ if $\xi = 1$ and $T = \infty$ if $\xi = 0$. Under the mixture formulation, $S(t)$ can be written as 
\begin{equation}
\label{mixture}
S(t) = S_{a}(t) (1-\eta) + \eta, 
\end{equation}
where  $\eta = \text{Pr}(\xi = 0)$ represents the cure fraction, and $S_{a}(t) = \text{Pr}(T > t | \xi = 1)$ denotes the survival function for susceptible individuals, also known as the latency survival function. Under right censoring, observations that are temporarily censored may become mixed with cured ones (long-term survivors), which can impact the identifiability of $\eta$ and $S_{a}(t)$. Let $C$ be the censoring time with the survival function $G(t) = \mbox{Pr}(C > t)$. 
Assume that $C$ is proper, with $G(\infty) = 0$, and that $T$ and $C$ are independent and do not experience simultaneous jumps. Observed variables include $X = \min(T, C)$ and $\delta = I(T \leq C)$, where  $I(A)$ is the indicator function that equals 1 if the event $A$ is true and 0 otherwise.  Denote $\zeta_C$ and $\zeta$ as the right end points of the supports of $C$ and $T|\xi = 1$, respectively. The condition of sufficient follow-up is related to the requirement that $\zeta \leq \zeta_C$, which means that the duration of follow-up is long enough to observe the largest event time  among susceptible individuals \cite{maller1996survival,maller1992estimating}.

\subsection{Nonparametric Analysis of the Cure Mixture Framework: A Review}
Denote $(T_i,C_i,\xi_i) \ (i = 1,\ldots,n)$ as identically and independently  distributed replications of $(T,C,\xi)$.
Observed variables include $(X_i, \delta_i)$, where $X_i = T_i \wedge C_i$ and $\delta_i = I(T_i \leq C_i)$ $(i = 1, \dots, n)$. Let $0 < t_{(1)} < \dots < t_{(K)}$ be the  distinct ordered failure times, $t_{(0)} = 0$ and $K$ be the number of distinct failure points.  
The KM estimator of $S(t)$ can be written as the product-limit form:
\begin{equation}
\label{product-limit_KM}
\hat{S}(t) = \prod_{k: t_{(k)} \leq t} \left(1 - \frac{d_{(k)}}{y_{(k)}}\right),
\end{equation}
where $d_{(k)} = \sum_{i=1}^n I(X_i = t_{(k)}, \delta_i = 1)$ and $y_{(k)} = \sum_{i=1}^n I(X_i \geq t_{(k)})$ $ (k = 1, \dots, K)$. 
The KM estimator possesses several desirable properties. 
It has been demonstrated that $\hat{F}(t) = 1- \hat{S}(t)$ can also be expressed as an IPCW estimator \cite{satten2001kaplan} such that 
\begin{equation}
\label{IPCWKM}
\hat{F}(t) = \sum_{i=1}^n \frac{I(X_i \leq t, \delta_i = 1)}{n\hat{G}(X_i)} = \frac{1}{n} \sum_{k: t_{(k)} \leq t} \tilde{d}_{(k)}, 
\end{equation}
where  $\tilde{d}_{(k)} = d_{(k)}/\hat{G}(t_{(k)})$ and $\hat{G}(t)$ is the KM estimator of $G(t)$ given by
\begin{equation*}
\hat{G}(t) = \prod_{k:u \leq t} \left(1 - \frac{\sum_{i=1}^n I(X_i = u, \delta_i = 0)}{\sum_{i=1}^n I(X_i \geq u)}\right).
\end{equation*} 
The mass assigned to $t_{(k)}$ is given by
\begin{equation*}
\hat{S}(\Delta t_{(k)}) \equiv \hat{S}(t_{(k-1)}) - \hat{S}(t_{(k)}) = \frac{d_{(k)}}{n\hat{G}(t_{(k)})} =  \frac{1}{n} \tilde{d}_{(k)},
\end{equation*}
where $\hat{S}(t_{(0)}) = \hat{S}(0) =1 $. 
When the largest observation is a censored observation, $ \hat{S}(t_{(K)})$ remains greater than 0, indicating that the KM curve reaches a plateau or levels off.  

Under sufficient follow-up with $\zeta \leq \zeta_C$,  the cure fraction $\eta$ can be estimated by
\begin{equation} 
\label{etahat}
\hat{\eta} = \hat{S}(t_{(K)})= 1-  \frac{\sum_{k=1}^K \tilde{d}_{(k)}}{n}.
\end{equation}  
Maller and Zhou established the properties of $\hat{\eta}$, such as consistency and asymptotic normality, given specific regularity conditions \cite{maller1996survival, maller1992estimating}. 
Under insufficient follow-up, Escobar-Bach and Van Keilegom \cite{EscobarBach2019} propose a new estimator under the assumption that $T|\xi = 1$ belongs to the maximum domain of attraction of an extreme value
 distribution. To estimate cure rates, they suggest utilizing the tail of the KM estimator, enhanced with a compensating term derived from extrapolation techniques in extreme value theory. Their formula is given by:
\begin{equation}
\label{etacheck}
    \check{\eta}_b = \hat{\eta} - \frac{\hat{S}(b t_{(K)}) - \hat{S}(t_{(K)})}{\check{b}_\gamma - 1},
\end{equation}
where \( b \in (0,1) \) is a scaling factor,  \( \gamma \) is an extreme value index that controls the tail behavior of \( S_a(t) \) and
\[
\check{b}_\gamma = \frac{\hat{S}(b t_{(K)}) - \hat{S}(b^2 t_{(K)})}{\hat{S}(t_{(K)}) - \hat{S}(b t_{(K)})}.
\]
Consequently, the survival function can be estimated by directly substituting an estimator of \(\eta\), either \(\hat{\eta}\) or \(\check{\eta}_b\), into equation (\ref{mixture}). The referenced paper suggests using bootstrap resampling to select the value of \(b\) \cite{EscobarBach2019}. 

\subsection{The Susceptible Survival Estimator under Sufficient Follow-Up}
Under sufficient follow-up, substituting $\hat{\eta}$ for $\eta$ in equation~(\ref{etahat}) and applying it to equation~(\ref{mixture}) yields the estimator:
\begin{equation}
\label{location-shift}
\hat{S}^{LS}_{a}(t) = \frac{\hat{S}(t) - \hat{\eta}}{1- \hat{\eta}},
\end{equation}
which can be interpreted as a location-scale-shift variant of \(\hat{S}(t)\).
We now demonstrate that $\hat{S}^{LS}_{a}(t)$ in \eqref{location-shift} is equivalent to the alternative IPCW and product-limit estimators:
\begin{equation}
\label{IPCW}
 \hat{S}^{W}_{a}(t)= \sum_{i=1}^n \frac{I(X_i > t, \delta_i = 1)}{n(1-\hat{\eta})\hat{G}(X_i)} = 
  \frac{1}{\hat{n}_{a}}  
  \sum_{k: t_{(k)} >t}\tilde{d}_{(k)}
  = \frac{\tilde{y}_{(k)}}{\hat{n}_{a}} 
\end{equation}
and
\begin{equation}
\label{product-limit}
\hat{S}_{a}^{PL} (t) = \prod_{k: t_{(k)} \leq t} \left(1 - \frac{\tilde{d}_{(k)}}{\tilde{y}_{(k)}}\right),
\end{equation}
where $\hat{n}_{a} = n(1-\hat{\eta})$, and $\tilde{y}_{(k)}$ is the adjusted number at risk at $t_{(k)}$ given by
\begin{equation*}
\tilde{y}_{(k)}=\sum_{i=1}^n \frac{I(X_i \geq t_{(k)},\delta_i=1)}{\hat{G}(X_i-)} = \sum_{j=k}^K\tilde{d}_{(k)}.
\end{equation*}
Note that $\tilde{y}_{(1)} = \sum_{k=1}^K\tilde{d}_{(k)}= \hat{n}_{a}$.  The expression $\hat{S}_{a}^W(t)$ in \eqref{IPCW} is an  IPCW  estimator of $S_{a}(t)$, with the mass assigned to each observed failure point adjusted by the estimated sample size $\hat{n}_{a}$ for the susceptible group. In addition, $\hat{S}_{a}^{PL} (t)$ in \eqref{product-limit} is an adjusted variant of the product-limit estimator in \eqref{product-limit_KM}.

Based on mass calculations, it can be shown that the three expressions in \eqref{location-shift}-\eqref{product-limit} are identical. Specifically, from the product-limit expression in \eqref{product-limit}, the mass at $t_{(k+1)}$ is given by:
\begin{equation*}
\hat{S}_{a}^{PL}(\Delta t_{(k+1)}) = \hat{S}_{a}^{PL} (t_{(k)})  \frac{\tilde{d}_{(k)}}{\tilde{y}_{(k)}}.
\end{equation*}
Since $\hat{S}^{PL}_{a}(t_{(0)}) =\hat{S}^{PL}_{a}(0) =1$, $\hat{S}^{PL}_{a}(\Delta t_{(1)}) = \tilde{d}_{(1)}/\tilde{y}_{(1)}$. It is straightforward to show that $\hat{S}^{PL}_{a}(t), \hat{S}^W_{a}(t)$ and $\hat{S}^{LS}_{a}(t)$ are equal and can be written as
\begin{equation}
\label{proposed}
\hat{S}_{a}(t) = \sum_{i=1}^{n} \frac{ I(X_i > t, \delta_i = 1)}{n(1-\hat{\eta})\hat{G}(X_i)} 
= \frac{1}{\hat{n}_{a}} \sum_{k: t_{(k)}>t} \tilde{d}_{(k)} = \frac{\tilde{y}_{(k)}}{\hat{n}_{a}}.
\end{equation}
We observe that the effect of long-term survivors on the estimation of $S_{a}(t)$ simply involves reducing the sample size to the susceptible subgroup. Consider the following three risk sets:
\begin{align*}
R(t) &= \{i: X_i \geq t \mid i=1,\ldots,n\}, \\
R_{a}(t) &= \{i: X_i \geq t, \xi_i=1 \mid i=1,\ldots,n\}, \\
\widetilde{R}(t) &= \{i: X_i \geq t, \delta_i=1 \mid i=1,\ldots,n\}.
\end{align*}
When there are long-term survivors with $\xi_i=0$, not all members in $R(t)$ are susceptible to the event of interest. However, in the presence of censoring, the susceptible risk set $R_{a}(t)$ is not fully recoverable from the available data. Given that $\widetilde{R}(t) \subset R_{a}(t) \subset R(t)$, $\widetilde{R}(t)$ can be used as a proxy for $R_{a}(t)$. To account for selection bias among observations in $\widetilde{R}(t)$, the inverse-probability-of-censoring weighting technique is employed to estimate the hazard rate of $T|\xi = 1$ at time $t_{(k)}$ as ${\widetilde{d}}_{(k)}/{\widetilde{y}}_{(k)}$, which explains \eqref{product-limit}. 

The theoretical properties of $\hat{S}_a(t)$  outlined in this subsection are instrumental for extending the analysis to different censoring scenarios, such as interval censoring, and offer a rationale for employing bootstrap re-sampling techniques for additional inferential objectives.
In a remarkable article, Gill \cite{gill1989non} showed that the nonparametric maximum likelihood estimator  is often determined as the solution of the likelihood equations for a collection of smooth parametric submodels. These equations are, in fact, precisely the ``self-consistency'' equations introduced by Efron \cite{ efron1967two}. Consequently, in many settings, a solution of the self-consistency equation is equivalent to the nonparametric maximum likelihood estimator, and in general settings, the nonparametric maximum likelihood estimator will be asymptotically efficient. 
Following the approach of Strawderman and Baer \cite{strawderman2023solutions}, we derive the self-consistency property for 
$\hat{S}^{PL}_{a}(t)$ in the cure model setting. The proposed self-consistency equation is a data-dependent function that represents the expected number of observations which are both susceptible and surviving at each time point in the sample. This function is expressed in terms of the target function $S_{a}(t)$  and specific nonparametric estimates. The representation of the self-consistency function, along with its proof, is provided in Section 1 of the Supplementary Materials. The adaptability of the self-consistency property to interval-censored data and more complex censoring schemes presents a potential avenue for future extensions within the cure mixture framework
\cite{Yu2000ConsistencySelfConsistent}.

We further study the weak convergence and nonparametric efficiency of the susceptible survival function. The weak convergence property of $\hat{S}_a(t)$ is given in  Theorem \ref{theorem1} and the property of nonparametrically efficiency is stated in  Theorem \ref{theorem_RAL}. 
\begin{theorem}
    Assume that $0 \le \eta < 1$,
    \begin{equation*}
        \sup_{0\le t < \infty}\bigg|\frac{Y(t)}{n} - S(t)G(t)\biggr|=o_P(1)
    \end{equation*} and
    \begin{equation}
    \label{regularity1}
        \int_0^{\zeta}\frac{dF(u)}{G(u-)}<\infty.
    \end{equation}
    Under sufficient follow-up with $\zeta \leq \zeta_C$, $\sqrt{n}(\hat{S}_a(t) - S_a(t))$ converges weakly to a mean-zero Gaussian process with $t\in[0,\zeta]$.
    \label{theorem1}
\end{theorem}
\noindent 
Note that when $\zeta < \zeta_C$, the condition in equation \eqref{regularity1} is satisfied. Therefore, it is necessary to verify condition \eqref{regularity1} only when $\zeta=\zeta_C$.  The sufficiency of the condition in \eqref{regularity1} is discussed in detail by Gill \cite{gill1983}.

The theory of regular and asymptotically linear (RAL) estimators can be applied to analyze $\hat{S}_a(t)$. This estimator is represented in terms of the efficient influence function, a concept introduced in the referenced book \cite{bickel1993efficient}. 
The influence function of the RAL estimator, which possesses the lowest asymptotic variance (van der Vaart, 1998, Theorem 25.20), corresponds to that of the asymptotically efficient estimator \cite{vaart_1998}.

Recall that the estimators $\hat{S}^{PL}_{a}(t), \hat{S}^{W}_{a}(t)$, and $\hat{S}^{LS}_{a}(t)$ all equal $\hat{S}_{a}(t)$ in (\ref{proposed}).
The efficient influence function form of a survival function estimate for general right-censored data is detailed in Section 4 of the book by Robins and Rotnitzky \cite{Robins1992} and on page 133 of the book by Van der Laan and Robins \cite{LaanRobins2003}.
Specializing those results for the model in (\ref{mixture}) and the estimation of $\eta$ gives the efficient influence function of $\hat{S}_{a}(t)$.

\begin{theorem} 
\label{theorem_RAL}
    Under sufficient follow-up with $\zeta \leq \zeta_C$, $\hat{S}_a(t)$ is regular and asymptotically linear (RAL) with the influence function
\begin{equation}
\psi_t(X, \delta) = \frac{\delta \left\{I(X>t) - S_a(t)\right\}}
{(1-\eta)G(X)}  + \int_0^{\zeta} \frac{q_t(u)}{\pi(u)} dM_{C,1}(u),
\end{equation}
where $\pi\left(u\right)$ and $q_t(u)$ are the probability limits of $Y\left(u\right)/n$ and 
\begin{equation*}
\sum_{i=1}^{n} \frac{\delta_iI(X_i\geq u) [I(X_i>t) - S_a(t)]}{n(1-\eta)G(X_i)},
\end{equation*}
respectively, $\lambda_c(t)$ is the hazard function of $C$, and the censoring martingale is
\begin{equation*}
M_{C,i}(t) =  I(X_i \leq t, \delta_i = 0) - \int_0^t I(X_i \geq u) \lambda_c(u)du.
\end{equation*}
It follows that $\hat{S}_a(t)$ is nonparametrically efficient and $\sqrt{n}(\hat{S}_a(t) - S_a(t))$ converges in distribution pointwise to a mean-zero normal random variable for $t\in[0,\zeta]$.
\end{theorem}

The representation in Theorem~\ref{theorem_RAL} is reminiscent of the representation of the KM estimator process as an identically and independently distributed process, as given by Lo and Singh \cite{lo1986product}. This representation justifies the bootstrap method for estimating the standard error of functionals of the  susceptible survival function estimate and its quantiles. Based on the bootstrap resampling approach, it provides a way of constructing confidence intervals (bands) for the unknown parameters (functionals of the distribution or quantile function). The proofs of the two theorems above are provided in Subsections 2.2 and 2.3 of the Supplementary Material. 

\subsection{The Susceptible Survival Estimator under insufficient Follow-Up}
Under conditions of insufficient follow-up, by substituting $\check{\eta}_b$ from equation~(\ref{etacheck}) into equation~(\ref{mixture}), we obtain:
\begin{equation}
\label{location-shift_check}
\check{S}_a^{LS}(t;b) = \frac{\hat{S}(t) - \check{\eta}_b}{1 - \check{\eta}_b}.
\end{equation}
The properties of \( \check{S}_a^{LS}(t;b) \) are determined by those of \( \hat{S}(t) \) and \( \check{\eta}_b \). For detailed properties of \( \check{\eta}_b \), refer to Escobar-Bach and Van Keilegom \cite{EscobarBach2019}. Although \( \check{\eta}_b \) does not ensure convergence to the true cure rate \( \eta \), it helps mitigate the underestimation typically associated with \( \hat{\eta} \) under insufficient follow-up.
As the duration of follow-up increases and approaches $\zeta$, the estimator $\check{\eta}_b$ becomes more accurate in approximating the true cure rate \( \eta \). This improvement in the accuracy of \(\check{\eta}_b\) directly enhances the performance of \(\check{S}_a^{LS}(t;b)\), making it more reliable in estimating the survival function \(S_a(t)\). In the supplementary materials, we offer heuristic discussions on the properties of \(\check{S}_a^{LS}(t;b)\), with a particular focus on analyzing its bias relative to \(S_a(t)\).

\section{Two-sample Application} 
\label{sec3}

We propose a flexible alternative by adapting the methodology of Tai et al. \cite{tai2023twosample}, which does not rely on specific model assumptions such as proportional hazards. Specifically, the tau process, which measures the relative performance of two groups under comparison, is defined as follows:
\begin{equation}
\label{tau}
\tau(t) = \int_{0}^{t} S_1(u) dF_0(u) - \int_{0}^{t} S_0(u) dF_1(u).
\end{equation}
Each component of the integrand of $\tau(t)$ represents the process where, at each failure time in one group, the survival probability of the other group is evaluated at that specific time point. The function $\tau(t)$ sums these differences in survival probabilities up to time $t$, providing a cumulative measure of the disparity between the two groups over time.  A positive value of $\tau(t)$ indicates that the treatment group (Group 1) exhibits a better effect up to time $t$. Note that $\tau(t)$ is a unitless and model-free measure, which makes it a robust and interpretable treatment-effect estimand, especially in the presence of non-proportional hazards \cite{tai2023twosample}. Let $\hat{\tau}(t)$ denote the estimator of $\tau(t)$. 

In the absence of a cure, such that $\mbox{Pr}(T < \infty) = 1$, $\tau(\infty)$ represents Kendall's tau correlation between the group indicator and the failure time. When cure is a possibility, the cross-comparison via the difference in the integrands of $\tau(t)$ is blurred by the presence of long-term survivors, as $S_1(u)$ and $S_0(u)$ may plateau at different levels. Specifically, when Group 1 exhibits a much higher cure rate than Group 0, leading to $S_1(u) \gg S_0(u)$ at large values of $u$, the differences in early stages may become obscured. In the next subsection, we define another tau process to compare event times between the two susceptible subgroups. We then explore its relationship with $\tau(t)$, as well as the cure rates $\eta_0$ and $\eta_1$.

\subsection{Tau Process for Susceptible Subgroups}

Under the two-sample setting, let $(T_\ell,C_\ell,\xi_\ell)$ be the failure time, censoring time and the indicator of susceptibility for Group $\ell$ with $\ell=0,1$. The mixture model described in \eqref{mixture} is adapted for each subgroup with $\xi_{\ell} = 1$, such that
\begin{equation}
\label{mixture1}
 S_\ell(t)  = S_{a,\ell}(t) (1-\eta_\ell) + \eta_\ell,  
\end{equation}
where  $S_{\ell}(t) = \mbox{Pr}(T_\ell > t)$,  $S_{a,\ell}(t) = \mbox{Pr}(T_\ell > t | \xi_\ell = 1)$  and $\eta_{\ell} = \mbox{Pr}(\xi_\ell = 0)$ for $\ell = 0,1$. Define $F_{a,\ell}(t) = 1- S_{a,\ell}(t)$ as the distribution function of $T_j|\xi_j = 1$ $(j = 0,1)$. The long-term treatment effect can be described by the difference of $\eta_1$ and $\eta_0$. The treatment effect on the susceptible groups is evaluated by quantifying the difference between $S_{a,1}(t)$ and $S_{a,0}(t)$. 

To characterize the treatment effect on the two susceptible groups, we introduce the susceptible tau process as follows:
\begin{equation}
\label{tau_a}
\tau_a(t)=\int_{0}^{t}{S_{a,1}(u)dF_{a,0}(u)}-\int_{0}^{t}{S_{a,0}(u)dF_{a,1}(u)}.    
\end{equation}
For susceptible patients who do not ultimately achieve a cure, a positive value of \(\tau_a(t)\) suggests that the treatment may still prolong the time until the occurrence of the unfavorable event of interest. Given that $\tau(t) = \Pr(T_0 < T_1 \wedge t) - \Pr(T_1 < T_0 \wedge t)$, 
we may think of \(\tau_a(t)\) as follows:
\begin{equation}
    \tau_a(t) = \frac{1}{(1-\eta_0)(1-\eta_1)} \left[ \left(\Pr(T_0 < T_1 \wedge t) - \eta_1 F_0(t)\right) - \left(\Pr(T_1 < T_0 \wedge t) - \eta_0 F_1(t)\right) \right],
\end{equation}
where \(-\eta_1 F_0(t)\) reflects the adjusted probability, excluding comparisons between susceptible subjects who die before \(t\) in group 0 and long-term survivors in group 1;  \(-\eta_0 F_1(t)\) reflects the adjusted probability, excluding comparisons between susceptible subjects who die before \(t\) in group 1 and long-term survivors in group 0; and $\left\{(1-\eta_0)(1-\eta_1)\right\}^{-1}$ serves as a normalizing constant.

Note that the signs of $\tau(t)$ and $\tau_a(t)$ may differ, indicating a potential reversal in effects between susceptible individuals and those considered long-term survivors. Specifically, we can write 
\begin{equation}
\label{tau_decomposition_1}
\tau\left(t\right)=\left(1-\eta_0\right)\left(1-\eta_1\right)\tau_a\left(t\right)+\left(1-\eta_0\right)\eta_1F_{a,0}\left(t\right)-\left(1-\eta_1\right)\eta_0F_{a,1}\left(t\right).    
\end{equation}
Consider a situation that $\eta_1 - \eta_0$ is far greater than zero but $S_{a,1}(t) < S_{a,0}(t)$ (equivalently, $F_{a,1}(t) > F_{a,0}(t)$). Equation~(\ref{tau_decomposition_1}) indicates that it is still possible for \(\tau(t) > 0\) while \(\tau_a(t) < 0\). In certain oncology clinical trials, this scenario may correspond to a situation where a subset of patients in Group 1 does not respond favorably to immunotherapy, resulting in their classification as short-term survivors, with even shorter survival. 

It has been argued that high-risk individuals are more likely to be depleted over time, which can make the hazard ratio susceptible to selection bias \cite{stensrud2019limitations,bartlett2020hazards}. Additionally, differences in cure rates may not fully capture treatment effects in the early stages of a trial, which are crucial for assessing overall efficacy. Therefore, to provide a comprehensive assessment of treatment effects, we recommend using multiple measures, including \(\tau(t)\), \(\tau_a(t)\), and \(\eta_1 - \eta_0\).

\subsection{Estimation of $\tau_a(t)$}
Observed data in the two-sample setting can be denoted as $(X_{\ell,i}, \delta_{\ell,i})$ for  $i = 1, \dots, n_\ell$ and $\ell=0,1$.
Let $\tilde{X}_{ij} = X_{0,i} \wedge  X_{1,j}$ and ${\widetilde{O}}_{ij}=I(X_{0,i}<X_{1,j},\delta_{0,i}=1)+I(X_{0,i}>X_{1,j},\delta_{1,j}=1)$. Under sufficient follow-up, $\hat{\eta}_\ell$ is a legitimate estimator of $\eta_{\ell}$ for $\ell = 0,1$.  The resulting estimator of $\tau_a(t)$ is given by 
\begin{equation}
\label{est_tau_a}
\hat{\tau}_a(t) = \frac{\sum_{i,j}\hat{\psi}_{ij}(t)\hat{w}_{ij}}{n_0 n_1(1-\hat{\eta}_0)(1-\hat{\eta}_1)},
\end{equation}
where
\begin{equation*}
\hat{\psi}_{ij}(t) = \frac{\tilde{O}_{ij}\text{sign}(X_{1,j}-X_{0,i})I(\widetilde{X}_{ij}\leq t)}{\hat{G}_0(\widetilde{X}_{ij})\hat{G}_1(\widetilde{X}_{ij})},
\end{equation*}
and
\begin{equation*}
\hat{w}_{ij} = \left[\frac{(1 - \hat{\eta}_0)\hat{S}_{a,0}(X_{0,i})}{(1 - \hat{\eta}_0)\hat{S}_{a,0}(X_{0,i}) + \hat{\eta}_0}\right]^{(1 - \delta_{0,i})} \times \left[\frac{(1 - \hat{\eta}_1)\hat{S}_{a,1}(X_{1,j})}{(1 - \hat{\eta}_1)\hat{S}_{a,1}(X_{1,j}) + \hat{\eta}_1}\right]^{(1 - \delta_{1,j})}.
\end{equation*}
The asymptotic property of $\hat{\tau}_a(t)$ is stated in Theorem~\ref{property_taua}, and its asymptotic variance can be obtained using the bootstrap approach, with the proof provided in Subsection 2.3 of the Supplementary Material.
 
\begin{theorem}
\label{property_taua}
    Suppose that the conditions of Theorem~\ref{theorem_RAL} hold for both groups, and let $n_1/n$ converge to $p_1$ as $n_0$ and $n_1$ tend to infinity, where $0 < p_1 < 1$. If $E[\psi_{ij}(t)w_{ij}]^2$ is finite for $t$ up to $\min(\zeta_0, \zeta_1)$, then $\sqrt{n}(\hat{\tau}_a(t) - \tau_a(t))$ converges pointwise to a mean-zero normal random variable.
\end{theorem}
\noindent 
One can apply a bootstrap approach for right censored data to develop inferential procedures.   The validity of such bootstrap confidence interval and tests follow using classical arguments for right censored data \cite{lo1986product, li1996, strawderman1997accurate}. The processes \(\hat{\tau}(t)\) and \(\hat{\tau}_a(t)\) can be implemented using the R package 'tauProcess' \cite{tauProcess}.

Under insufficient follow-up, the estimator of $\eta_{\ell}$ can be obtained by modifying Equation~(\ref{etacheck}) for the two-sample setting, and is denoted as $\check{\eta}_{\ell, b_{\ell}}$ for $\ell = 0, 1$.
The corresponding estimator of $\tau_a(t)$ can be modified by replacing $\hat{\eta}_\ell$ with $\check{\eta}_{\ell, b_\ell}$, and is denoted as $\check{\tau}_a(t; b_0, b_1)$.
Note that the tail behaviors of the two groups may differ, necessitating the separate estimation of $b_0$ and $b_1$ in practical applications. The implementation procedure suggested by Escobar-Bach and Van Keilegom will be summarized in the data analysis section \cite{EscobarBach2019}.

\section{Simulation Study} 
In the first design, we assess the performance of the estimators for \(S_a(t)\) and \(\eta\) under both sufficient and insufficient follow-up conditions. In the second simulation design, we evaluate the estimators of \(\tau_a(t)\) for two scenarios—crossing and non-crossing survival functions \(S_{a,0}(t)\) and \(S_{a,1}(t)\), with \(\eta_\ell\) values of 0.2 and 0.4 for \(\ell = 0, 1\). For both simulation settings, we analyze the average bias of each estimator, along with the bootstrap standard deviation estimates and the empirical coverage probabilities of the estimated confidence intervals. These results are derived using 2000 bootstrap resamples across 500 simulation runs. Additional simulation results are detailed in the Supplementary Material.

\subsection{Finite-Sample Performance of Estimators for $S_a(t)$ and $\eta$}
\label{subsection:estimator-Sa}
In the initial design, we assess the finite-sample performance of estimators for $S_a(t)$ and $\eta$, setting $\eta = 0.2$ and $0.4$. To model $T|\xi = 1$ with bounded support, we employ a Beta distribution with parameters $\alpha_1 = 1$ and $\alpha_2 = 3$.
We first discuss the results under conditions where $\zeta = \zeta_C$ and \eqref{regularity1} is met, with the censoring variable $C$ following a $\text{Uniform}[0, 4]$ distribution. As shown in Table~\ref{tab:Sa_Estimation_Sufficient}, the average biases of $\hat{S}_a(t)$ and $\hat{\eta}$ are close to zero, and the bootstrap estimates of the standard deviation for $\hat{S}_a(t)$ closely match the empirical estimates. 
The empirical coverage probabilities are also around the 95\% nominal level. Note that confidence intervals are wider under $\eta = 0.4$ compared to $\eta = 0.2$. Additional analyses based on other scenarios are presented in the Supplementary materials.

\begin{center}
[Insert Table~\ref{tab:Sa_Estimation_Sufficient} \& Table~\ref{tab:Sa_Estimation_Insufficient}]
\end{center}

Under insufficient follow-up,  the censoring variable $C$ following a $\text{Uniform}[0, 0.8]$ distribution.
We evaluate the performance of \(\check{S}_a^{LS}(t;b_*)\) using \(\check{\eta}_{b_*}\) to estimate \(\eta\), where \(b_*\) is selected such that the corresponding estimator \(\check{\eta}_b\) closely matches the average outcome of a bootstrap experiment \cite{EscobarBach2019}. This choice minimizes the deviation between \(\check{\eta}_b\) and the average bootstrap estimator across multiple resampled datasets.
Compared with the performance of \(\hat{\eta}\) shown in Table~\ref{tab:Sa_Estimation_Sufficient}, \(\check{S}_a^{LS}(t;b_*)\) exhibits slight bias, as indicated in row \((a)\) of the last column in Table~\ref{tab:Sa_Estimation_Insufficient}.
Nevertheless, the performance of \(\check{S}_a^{LS}(t;b_*)\) remains satisfactory as an estimator of \(S_a(t)\), although the coverage probability tends to deviate more from 95\% as \(t\) increases.

\subsection{Finite-Sample Performance of $\hat{\tau}_a(t)$}
In this subsection, we present results for \(\hat{\tau}_a(t)\) under conditions of sufficient follow-up. Note that under insufficient follow-up, the performance of \(\check{\tau}_a(t; b_0, b_1)\) is influenced by the shapes of the unobserved tail distributions. Therefore, we apply this modified estimator in our data analysis but exclude it from the simulations.
Figures S4 and S5 in the Supplementary material depict  two non-crossing susceptible survival functions with the same cure rates and the corresponding $\tau_a(t)$. We present the results for the case with apparent disparity, and the other case is provided in the Supplementary material.
From Table~\ref{tau_no_crossing_2},  
we observe that the average bias of $\hat{\tau}_a(t)$ is almost zero. The bootstrap estimates of the standard deviation of $\hat{\tau}_a(t)$ closely align with the empirical estimates, and the empirical coverage probabilities are close to the 95\% nominal level.  
Notice that the lengths of the confidence intervals are associated with the values of $\eta_0$ and $\eta_1$. Higher cure rates lead to wider confidence intervals for $\tau_a(t)$. According to Table~\ref{tau_no_crossing_2}, $\hat{\tau}_a(1)$ yields a significant result for testing $H_0: \tau_a(1) = 0$.

\begin{center}
[Insert Table~\ref{tau_no_crossing_2}]
\end{center}

In the Supplementary Material, Figures S6 and S7 depict scenarios where the susceptible survival functions intersect at a susceptible survival probability of approximately 0.5. The former corresponds to $\eta_0=\eta_1 = 0.2$, and the latter corresponds to $\eta_0=0.2, \eta_1 = 0.4$. The corresponding simulation results are summarized in Table~\ref{tau_crossing_1}. The performance of $\hat{\tau}_a(t)$ is similar to that of the previous settings. The result from the last column indicates that there is no significant result for testing $H_0: \tau_a(1) = 0$ based on $\hat{\tau}_a(1)$.

\begin{center}
[Insert Table~\ref{tau_crossing_1}]
\end{center}

In Figure S8 of the Supplementary Material, we present simulation results that examine the situation where $\eta_1 > 0$ and $\eta_0 = 0$. These results confirm the validity of the inference procedure based on $\hat{\tau}_a(t)$.

\section{Data Analysis} 
The CheckMate 067 trial, a randomized, multicenter, phase 3 study conducted from July 2013 to March 2014, evaluated three different immunotherapy strategies for advanced melanoma in a cohort of 945 patients. Participants were assigned to receive either a combination of nivolumab and ipilimumab, nivolumab alone, or ipilimumab alone, with each regimen designed to boost the immune system's ability to combat the disease.

\begin{figure}[htbp]
\centering
\includegraphics[width=0.9\textwidth]{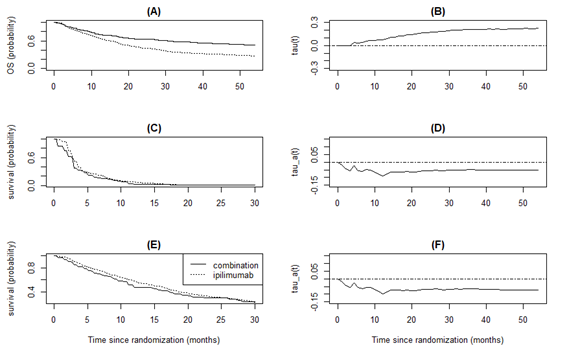}
\caption{Two-sample comparison between "nivolumab + ipilimumab" (Group 1) and "ipilimumab alone" (Group 0) based on digitized data from CheckMate 067 \cite{hodi2018nivolumab}. (A) Kaplan-Meier curves; (B) Plot of $\hat{\tau}(t)$; (C) Curves of $\hat{S}_{a,\ell}(t)$ for $\ell = 0,1$; (D) Plot of $\hat{\tau}_a(t)$; (E) Curves of  \( \check{S}_{a,\ell}^{LS}(t;b_{*,\ell}) \) for $\ell = 0,1$; (F) Plot of $\check{\tau}_a(t; b_{*,0}, b_{*,1})$.
}
\label{fig_modify2_combined}
\end{figure}

\begin{figure}[htbp]
\centering
\includegraphics[width=0.9\textwidth]{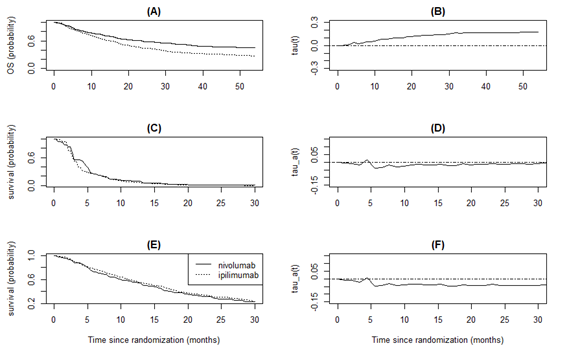}
\caption{Two-sample comparison between "nivolumab alone" (Group 1) and "ipilimumab alone" (Group 0) based on digitized data from CheckMate 067 \cite{hodi2018nivolumab}. (A) Kaplan-Meier curves; (B) Plot of $\hat{\tau}(t)$; (C) Curves of $\hat{S}_{a,\ell}(t)$ for $\ell = 0,1$; (D) Plot of $\hat{\tau}_a(t)$; (E) Curves of  \( \check{S}_{a,\ell}^{LS}(t;b_{*,\ell}) \) for $\ell = 0,1$; (F) Plot of $\check{\tau}_a(t; b_{*,0}, b_{*,1})$.
}
\label{fig_modify2_alone}
\end{figure}

First, we apply our proposed methodology by digitizing the overall survival KM curves for the three treatment groups based on the 4-year report \cite{hodi2018nivolumab}.
The estimated cure rates based on the tail values of the KM curves are 0.516 for the combination treatment, 0.452 for nivolumab alone, and 0.272 for ipilimumab alone. Given that the 4-year data may not represent sufficient follow-up, we also applied the extrapolation method proposed by Escobar-Bach and Van Keilegom \cite{EscobarBach2019}. This approach provided cure rate estimates based on $\check{\eta}_{b_*}$, yielding values of 0.481 for combination therapy, 0.421 for nivolumab alone, and 0.199 for ipilimumab alone. 
The difference in cure rates between the combined group and 'ipilimumab alone' is significant, calculated as $0.516 - 0.272 = 0.244\ (0.164, 0.323)$ with p-value $1.93\times 10^{-9}$ using the KM tail estimates, and $0.481 - 0.199 = 0.282\ (0.128, 0.445)$ with p-value $3.97\times 10^{-4}$ using the modified estimator. Figure~\ref{fig_modify2_combined} offers a graphical assessment, showcasing the KM curves, $\hat{\tau}(t)$, as well as both versions of the estimated susceptible survival functions and the susceptible tau process. In plot (A), the combined group shows a higher survival curve and exhibits a higher cure rate. Plot (B) shows that $\hat{\tau}(t)$ becomes positive starting from the third month onward.
Using $\hat{\eta}_{\ell}$ as the estimate of $\eta_{\ell}$, plots (C) and (D) display $\hat{S}_{a,\ell}(t)$ for $\ell = 0, 1$ and $\hat{\tau}_a(t)$, respectively.
Using $\check{\eta}_{\ell,b_{*,\ell}}$ as the estimate of $\eta_{\ell}$, plots (E) and (F) display $\check{S}_{a,\ell}^{LS}(t;b_{*,\ell})$ for $\ell = 0, 1$ and $\check{\tau}_a(t; b_{*,0}, b_{*,1})$, respectively. By comparing the last two rows of Figure~\ref{fig_modify2_combined}, we observe that the modified cure rate estimators do not significantly impact the estimation of the susceptible survival functions and the susceptible tau process.
Although the combined therapy clearly outperforms 'ipilimumab alone' based on the KM estimators and $\hat{\tau}(t)$, the analysis of susceptible groups indicates  reversal relationships. This suggests that while a larger proportion of patients in the combined therapy group are long-term survivors, those who do not achieve a durable effect have similar or slightly shorter survival times compared to their counterparts in Group 0.
The difference in cure rates between 'nivolumab alone' and 'ipilimumab alone' is significant, calculated as $0.452 - 0.272 = 0.18\ (0.101, 0.260)$ with p-value $9.08 \times 10^{-6}$ using the KM tail estimates, and $0.421 - 0.199 = 0.222\ (0.067, 0.384)$ with p-value $0.005$ using the modified estimator. According to Figure~\ref{fig_modify2_alone}, although the 'nivolumab alone' group exhibits a higher cure rate, the estimated susceptible survival functions are roughly similar. The estimated curves of $\tau_a(t)$ demonstrate their proximity to zero and even suggest a slight reversal, as observed in the third row.

\begin{center}
[Insert Figure~\ref{fig_modify2_combined} \& Figure~\ref{fig_modify2_alone}]
\end{center}


\section{Concluding Remarks} 
Long-term survivorship, often referred to as cure, is a common outcome in various fields. In the context of cancer treatment, the sustained effectiveness of immunotherapy and other advanced options has generated optimism among both patients and healthcare professionals. 
However, understanding the potential heterogeneity in patient responses to these treatments remains a crucial aspect of devising appropriate treatment strategies for the right individuals. Cure mixture models provide a useful framework that allows for separate evaluation of long-term survivors and susceptible individuals who do not achieve the desired long-term status.
While the estimation of the cure rate assumes sufficient follow-up, which may not always be attainable, extreme value theory provides methods to develop new estimators \cite{EscobarBach2019}. These estimators can reduce bias in the tail estimates of the KM curves under conditions of insufficient follow-up.
Under sufficient follow-up, we demonstrate that the estimator of the susceptible survival function, $\hat{S}_a(t)$, inherits many favorable properties of the KM estimator, which paves the way for further extensions across various data structures and settings.
Under insufficient follow-up, the location-scale-shift version adapts the modified cure rate estimator suggested by Escobar-Bach and Van Keilegom \cite{EscobarBach2019}. To evaluate the effect on long-term survivors, we recommend using the difference in cure rates. Additionally, our proposed graphical estimand, $\tau_a(t)$, offers insights into the treatment effects over time for individuals who have not been cured, highlighting the timing and impact of the therapy. The susceptible tau process can be estimated nonparametrically, provided that a suitable cure rate estimator is available.
Building on these methods, employing multiple estimands in clinical trial analysis provides a clear framework that enhances understanding of survival outcomes. This approach not only enriches the analysis but also informs future therapeutic strategies and research directions. 

\section*{Acknowledgements}
We express our gratitude to Robert Strawderman and Benjamin Baer for their invaluable discussions on self-consistency and efficiency theory. Additionally, we are thankful to Mikael Escobar-Bach and Ingrid Van Keilegom for kindly providing the code for their cure rate estimator.

\clearpage

\clearpage

\clearpage
\begin{table}[ht]
\centering
\small\sf
\caption{Finite-sample performance of $\hat{S}_a(t)$ and $\hat{\eta}$ based on $n=200$, when $S_a(t)$ follows a Beta(1,3) distribution and $C \sim U(0,4)$, with $\eta = 0.2$ resulting in $\text{Pr}(\delta = 0) \approx 0.401$ (upper) and $\eta = 0.4$ resulting in $\text{Pr}(\delta = 0) \approx 0.550$ (lower).}
\label{tab:Sa_Estimation_Sufficient}
\resizebox{0.8\textwidth}{!}{
\begin{tabular}{@{}llllllll@{}}
\toprule
$t$ & 0.091 & 0.134 & 0.181 & 0.234 & 0.295 & 0.370 & Cure Rate \\
\midrule
$S_a(t)$ & 0.75 & 0.65 & 0.55 & 0.45 & 0.35 & 0.25  & $\eta = 0.2$ \\
(a) & 0.001 & 0.001 & 0.001 & -0.001 & -0.001 & -0.001 & 0.002 \\
(b) & 0.036 & 0.041 & 0.045 & 0.047 & 0.047 & 0.046 & 0.042 \\
(c) & 0.035 & 0.041 & 0.044 & 0.046 & 0.048 & 0.047 & 0.044 \\
(d) & 0.946 & 0.948 & 0.943 & 0.948 & 0.940 & 0.934 & 0.944 \\
(e) & 0.143 & 0.162 & 0.176 & 0.184 & 0.186 & 0.181 & 0.165 \\
\midrule
$S_a(t)$ & 0.75 & 0.65 & 0.55 & 0.45 & 0.35 & 0.25  & $\eta = 0.4$ \\
(a) & 0.001 & 0.001 & 0.000 & -0.001 & -0.001 & -0.001 & 0.000 \\
(b) & 0.042 & 0.049 & 0.053 & 0.055 & 0.056 & 0.054 & 0.047 \\
(c) & 0.041 & 0.048 & 0.052 & 0.055 & 0.057 & 0.056 & 0.047 \\
(d) & 0.952 & 0.950 & 0.949 & 0.944 & 0.936 & 0.935 & 0.949 \\
(e) & 0.167 & 0.190 & 0.207 & 0.216 & 0.219 & 0.213 & 0.184 \\
\bottomrule
\end{tabular}
}
\newline
\textbf{Notes:} (a) Average bias; (b) Estimated standard deviation by bootstrap; (c) Empirical standard deviation; (d) Empirical coverage probability of the 95\% confidence interval based on (b); (e) Average length of the 95\% confidence interval.
\end{table}

\clearpage

\begin{table}[ht]
\centering
\small\sf
\caption{Finite-sample performance of $\check{S}_a(t;b_*)$ and $\check{\eta}_{b_*}$ based on $n=200$, when $S_a(t)$ follows a Beta(1,3) distribution and $C \sim U(0,0.8)$, with $\eta = 0.2$ resulting in $\text{Pr}(\delta = 0) \approx 0.448$ (upper) and $\eta = 0.4$ resulting in $\text{Pr}(\delta = 0) \approx 0.584$ (lower).}
\label{tab:Sa_Estimation_Insufficient}
\resizebox{0.8\textwidth}{!}{
\begin{tabular}{@{}llllllll@{}}
\toprule
$t$ & 0.091 & 0.134 & 0.181 & 0.234 & 0.295 & 0.370 & Cure Rate \\
\midrule
$S_a(t)$ & 0.75 & 0.65 & 0.55 & 0.45 & 0.35 & 0.25  & $\eta = 0.2$ \\
(a) & -0.003 & -0.004 & -0.004 & -0.004 & -0.007 & -0.009 & -0.052 \\
(b) & 0.038 & 0.044 & 0.048 & 0.052 & 0.053 & 0.053 & 0.068 \\
(c) & 0.037 & 0.044 & 0.049 & 0.053 & 0.056 & 0.057 & 0.071 \\
(d) & 0.956 & 0.954 & 0.936 & 0.938 & 0.922 & 0.908 & 0.868 \\
(e) & 0.150 & 0.173 & 0.190 & 0.202 & 0.210 & 0.210 & 0.267 \\
\midrule
$S_a(t)$ & 0.75 & 0.65 & 0.55 & 0.45 & 0.35 & 0.25  & $\eta = 0.4$ \\
(a) & -0.004 & -0.004 & -0.003 & -0.004 & -0.008 & -0.009 & -0.065 \\
(b) & 0.045 & 0.052 & 0.057 & 0.061 & 0.064 & 0.063 & 0.083 \\
(c) & 0.043 & 0.052 & 0.058 & 0.064 & 0.069 & 0.069 & 0.088 \\
(d) & 0.954 & 0.942 & 0.942 & 0.928 & 0.934 & 0.900 & 0.898 \\
(e) & 0.176 & 0.204 & 0.225 & 0.241 & 0.249 & 0.248 & 0.324 \\
\bottomrule
\end{tabular}
}
\newline
\textbf{Notes:} (a) Average bias; (b) Estimated standard deviation by bootstrap; (c) Empirical standard deviation; (d) Empirical coverage probability of the 95\% confidence interval based on (b); (e) Average length of the 95\% confidence interval.
\end{table}

\clearpage
\begin{table}[!t]
\small\sf\centering
\caption{Finite-sample Performance of $\hat{\tau}_a(t)$, based on $n_0=n_1=200$, with $T_1|\xi_1 = 1 \sim$ Beta(1,2), $T_0|\xi_0 = 1 \sim$ Beta(1,4), $\eta_0 = \eta_1 = 0.2$ (top) and $\eta_0 = \eta_1 = 0.4$ (bottom).}
\label{tau_no_crossing_2}
\resizebox{\textwidth}{!}{%
\begin{tabular}{@{\extracolsep{\fill}}lllllllllll@{}}
\toprule
& \multicolumn{10}{c}{$\eta_0 = \eta_1 = 0.2, \mbox{Pr}(\delta_0 = 0) = 0.363$, $\mbox{Pr}(\delta_1 = 0) = 0.464$}  \\
\midrule
$t$ & 0.1 & 0.2 & 0.3 & 0.4 & 0.5 & 0.6 & 0.7 & 0.8 & 0.9 & 1.0 \\
$\tau_a(t)$ & 0.156 & 0.246 & 0.294 & 0.318 & 0.328 & 0.332 & 0.333 & 0.333 & 0.333 & 0.333 \\
\midrule
(a) & 0.001 & -0.002 & -0.001 & -0.001 & -0.001 & -0.001 & -0.001 & -0.002 & -0.002 & -0.002 \\
(b) & 0.053 & 0.065 & 0.071 & 0.074 & 0.075 & 0.076 & 0.076 & 0.076 & 0.076 & 0.076 \\
(c) & 0.053 & 0.065 & 0.069 & 0.071 & 0.074 & 0.074 & 0.074 & 0.074 & 0.074 & 0.074 \\
(d) & 0.948 & 0.954 & 0.952 & 0.948 & 0.946 & 0.950 & 0.948 & 0.948 & 0.948 & 0.948 \\
(e) & 0.208 & 0.257 & 0.279 & 0.290 & 0.295 & 0.297 & 0.298 & 0.298 & 0.298 & 0.298 \\
\midrule
& \multicolumn{10}{c}{$\eta_0 = \eta_1 = 0.4, \mbox{Pr}(\delta_0 = 0) = 0.520$, $\mbox{Pr}(\delta_1 = 0) = 0.603$}  \\
\midrule
$t$ & 0.1 & 0.2 & 0.3 & 0.4 & 0.5 & 0.6 & 0.7 & 0.8 & 0.9 & 1.0 \\
$\tau_a(t)$ & 0.156 & 0.246 & 0.294 & 0.318 & 0.328 & 0.332 & 0.333 & 0.333 & 0.333 & 0.333 \\
\midrule
(a) & 0.003 & 0.000 & -0.003 & -0.003 & -0.003 & -0.003 & -0.004 & -0.004 & -0.004 & -0.004 \\
(b) & 0.062 & 0.078 & 0.085 & 0.089 & 0.090 & 0.091 & 0.091 & 0.091 & 0.091 & 0.091 \\
(c) & 0.059 & 0.076 & 0.084 & 0.088 & 0.089 & 0.090 & 0.090 & 0.091 & 0.091 & 0.091 \\
(d) & 0.954 & 0.958 & 0.950 & 0.954 & 0.954 & 0.950 & 0.950 & 0.950 & 0.950 & 0.950 \\
(e) & 0.243 & 0.304 & 0.333 & 0.348 & 0.354 & 0.357 & 0.358 & 0.358 & 0.358 & 0.358 \\
\bottomrule
\end{tabular}
}
\newline
\textbf{Notes:} (a) Average bias; (b) Estimated standard deviation by bootstrap; (c) Empirical standard deviation; (d) Empirical coverage probability of the 95\% confidence interval based on (b); (e) Average length of the 95\% confidence interval.
\end{table}

\clearpage

\begin{table}[!t]
\small\sf\centering
\caption{Finite-sample Performance of $\hat{\tau}_a(t)$, based on $n_0=n_1=200$, with $T_1|\xi_1 = 1 \sim$ Beta(0.5,1.5), $T_0|\xi_0 = 1 \sim$ Beta(1,4), $\eta_0 = \eta_1 = 0.2$ (top) and $\eta_0 = 0.2, \eta_1 = 0.4$ (bottom).
\label{tau_crossing_1}}
\resizebox{\textwidth}{!}{%
\begin{tabular}{@{\extracolsep{\fill}}lllllllllll@{}}
\toprule
& \multicolumn{10}{c}{$\eta_0 = \eta_1 = 0.2, \mbox{Pr}(\delta_0 = 0) = 0.358$, $\mbox{Pr}(\delta_1 = 0) = 0.401$} \\
\midrule
$t$ & 0.1 & 0.2 & 0.3 & 0.4 & 0.5 & 0.6 & 0.7 & 0.8 & 0.9 & 1.0 \\
$\tau_a(t)$ & -0.093 & -0.046 & -0.015 & 0.002 & 0.010 & 0.014 & 0.015 & 0.015 & 0.015 & 0.015 \\
\midrule
(a) & 0.000 & 0.001 & 0.000 & 0.000 & -0.001 & 0.000 & 0.000 & 0.000 & 0.000 & 0.000 \\
(b) & 0.062 & 0.074 & 0.080 & 0.083 & 0.085 & 0.086 & 0.086 & 0.086 & 0.086 & 0.086 \\
(c) & 0.061 & 0.074 & 0.082 & 0.085 & 0.087 & 0.088 & 0.088 & 0.088 & 0.088 & 0.088 \\
(d) & 0.954 & 0.954 & 0.944 & 0.946 & 0.946 & 0.942 & 0.944 & 0.944 & 0.942 & 0.942 \\
(e) & 0.244 & 0.290 & 0.314 & 0.327 & 0.333 & 0.336 & 0.337 & 0.338 & 0.338 & 0.338 \\
\midrule
& \multicolumn{10}{c}{$\eta_0 = 0.2, \eta_1 = 0.4, \mbox{Pr}(\delta_0 = 0) = 0.358$, $\mbox{Pr}(\delta_1 = 0) = 0.552$} \\
\midrule
$t$ & 0.1 & 0.2 & 0.3 & 0.4 & 0.5 & 0.6 & 0.7 & 0.8 & 0.9 & 1.0 \\
$\tau_a(t)$ & -0.093 & -0.046 & -0.015 & 0.002 & 0.010 & 0.014 & 0.015 & 0.015 & 0.015 & 0.015 \\
\midrule
(a) & 0.002 & 0.004 & 0.003 & 0.003 & 0.003 & 0.003 & 0.003 & 0.003 & 0.003 & 0.003 \\
(b) & 0.070 & 0.084 & 0.092 & 0.096 & 0.098 & 0.099 & 0.099 & 0.100 & 0.100 & 0.100 \\
(c) & 0.065 & 0.081 & 0.091 & 0.095 & 0.097 & 0.098 & 0.099 & 0.099 & 0.099 & 0.099 \\
(d) & 0.968 & 0.956 & 0.950 & 0.948 & 0.944 & 0.944 & 0.942 & 0.938 & 0.938 & 0.938 \\
(e) & 0.275 & 0.329 & 0.359 & 0.376 & 0.385 & 0.389 & 0.390 & 0.390 & 0.390 & 0.390 \\
\bottomrule
\end{tabular}
}
\newline
\textbf{Notes:} (a) Average bias; (b) Estimated standard deviation by bootstrap; (c) Empirical standard deviation; (d) Empirical coverage probability of the 95\% confidence interval based on (b); (e) Average length of the 95\% confidence interval.
\end{table}

\clearpage
\bibliographystyle{unsrt}
\bibliography{Cure_SMMR_reference_expanded}

\end{document}


\maketitle
\vspace{-2cm} 

\begin{center}
Yi-Cheng Tai\textsuperscript{1,2}, Weijing Wang\textsuperscript{2*}, Martin T. Wells\textsuperscript{1} \\
\textsuperscript{1}
Department of Statistics and Data Science, Cornell University, Ithaca, NY 14853, USA \\
\textsuperscript{2}
Institute of Statistics, National Yang Ming Chiao Tung University, Hsin Chu City, Taiwan, ROC \\
wjwang@stat.nycu.edu.tw\textsuperscript{*}
\end{center}

\setstretch{1.5} 

\section*{Introduction}

In this Supplementary Material, we provide additional details and proofs that complement our main manuscript, 'Nonparametric Inference Methods for Analyzing Survival Data with Long-Term Survivors.' The content is organized into several sections. Under conditions of sufficient follow-up, we develop the self-consistency equation for $\hat{S}_a(t)$ in Section~\ref{self_consistency}. Section~\ref{Proofs} presents the proofs for \emph{Theorem 2.1}, \emph{Theorem 2.2}, and \emph{Theorem 3.1}. In Section~\ref{Insufficient_Follow_Up}, we provide heuristic discussions on the properties of modified estimators developed for conditions of insufficient follow-up.
Section~\ref{simulations} includes additional simulation results for $\hat{S}_{a}(t)$ and $\hat{\tau}_{a}(t)$, discussions on the absence of cure in one group, and further data analysis.

\section{The Self-Consistency Equation for $\hat{S}_{a}(t)$}
\label{self_consistency}
\subsection{Technical Foundations}
\label{appendix_proof_self_consistency}
Applying ``self-consistency'' equations results in \cite{strawderman2023solutions} provides novel, non-inductive proof that the product-limit estimator for the survival function (the NPMLE) solves Efron's self-consistency equation. \cite{strawderman2023solutions} also show that the use of the Volterra integral equations \citep{gill1990survey} gives a simple and direct way to prove that the product-limit estimator has an inverse-probability-censoring-weighting representation. 
 We will apply the same ideas to derive a similar property for $\hat{S}^{PL}_{a}(t)$. 
 
Recall that 
\begin{align*}
R(t) &= \{i: X_i \geq t \mid i=1,\ldots,n\}, \\
R_{a}(t) &= \{i: X_i \geq t, \xi_i=1 \mid i=1,\ldots,n\}. 
\end{align*}
Given that $R_{a}(t)\subset R(t)$, define
\begin{equation}
\label{Phi}
\varphi(t) = \mbox{Pr}\left(i\in R_{a}(t)|i\in R(t)\right) 
= \frac{\mbox{Pr}(X_i\geq t,\xi_i=1)}{\mbox{Pr}(X_i\geq t)},
\end{equation}
which describes the proportion of susceptible individuals in the risk set $R(t)$ at time $t$. The function $\varphi(t) $ can be estimated by
\begin{equation}
\label{estPhi}
\hat{\varphi}(t) = 1-\frac{\hat{\eta}\hat{G}(t-)}{Y(t)/n}, \quad
\end{equation}
where $Y(t)=\sum_{i=1}^n I(X_i\geq t)$ is the number at risk at time $t$. 
For a censored observation $X_i=x_i\le t$ with $\delta_i=0$, the conditional probability that it would be susceptible and fail beyond time $t$ is given by
\begin{align*}
\mbox{Pr}(T_i>t,\xi_i=1|X_i=x_i,\delta_i=0) 
&= \mbox{Pr}\left(i\in R_{a}(x_i+)\mid i\in R(x_i+)\right)\frac{S_{a}(t)}{S_{a}(x_i)} \\
&= \varphi(x_i +)\frac{S_{a}(t)}{S_{a}(x_i)}.
\end{align*}
The expected number of susceptible individuals contained in $R(t)$ is $n H_{1,a}(t-)$, where 
\begin{equation*}
H_{1,a}(t) = \mbox{Pr}(X>t,\xi=1) = \mbox{Pr}(X>t) - \mbox{Pr}(C>t,\xi=0) = \mbox{Pr}(X>t) - \eta G(t),
\end{equation*}
which can be estimated by
\begin{equation*}
\hat{H}_{1,a}(t) = \frac{1}{n}\sum_{i=1}^n I(X_i>t) - \hat{\eta}\hat{G}(t) = \frac{1}{n}Y(t+) - \hat{\eta}\hat{G}(t). 
\end{equation*}
A self-consistent estimator of $S_{a}(t)$, denoted as $\widetilde{S}_{a}(t)$, satisfies the following equation: 
\begin{equation}
\label{self-consistency}
n(1-\hat{\eta})\widetilde{S}_{a}(t) = \sum_{i=1}^{n}I(X_i \leq t, \delta_i=0)\left\{\hat{\varphi}(X_i +)\frac{\widetilde{S}_{a}(t)}{\widetilde{S}_{a}(X_i)}\right\} + n\hat{H}_{1,a}(t).
\end{equation}

\begin{center}
[Insert Figure~\ref{fig_self_consistency}]
\end{center}

Figure~\ref{fig_self_consistency}  illustrates the main idea of the algorithm.
The bracket (a) in Figure~\ref{fig_self_consistency} explains the selection mechanism described by Equation \eqref{self-consistency}. In the bracket (b) of Figure \ref{fig_self_consistency}, the number of censored and cured observations is estimated and then excluded in the self-consistency equation. 
When there are no long-term survivors, $n{\hat{H}}_{1,a}\left(t\right)=\sum_{i=1}^{n}{I(X_i>t)}$ and  $\hat{\varphi}(t)=\varphi(t)=1$ for all $t$. In this case, equation \eqref{self-consistency} reduces to the original self-consistency equation: 
 \begin{align*}
n\hat{S}\left(t\right)&=\sum_{i=1}^{n}{I\left(X_i\le t,\delta_i=0\right)\frac{\hat{S}\left(t\right)}{\hat{S}\left(X_i\right)}}+\sum_{i=1}^{n}I\left(X_i>t\right).
\end{align*}

\subsection{Mathematical Proof}

We now prove that $\hat{S}^{PL}_{a}(t)$ is the unique solution to the self-consistency equation in \eqref{self-consistency}, using its representation as a Volterra integral equation.  As a corollary, $\hat{S}^{PL}_{a}(t)$ is the nonparametric maximum likelihood estimator. 
Our proof follows a similar line of reasoning as the arguments presented in \cite{strawderman2023solutions}. We can write $\hat{S}_{a}(t)$ as 
\begin{align*}
\hat{S}^{PL}_{a}(t) &= \prod_{0\leq u \leq t}\left\{1-\frac{dN(u)}{Y(u)-n\hat{\eta}\hat{G}(u-)}\right\} \\
&= \prod_{0\leq u \leq t}\left\{\frac{Y(u)-n\hat{\eta}\hat{G}(u-) - dN(u)}{Y(u)-n\hat{\eta}\hat{G}(u-)}\right\},
\end{align*}
where $N\left(t\right)=\sum_{i=1}^{n}I\left(X_i\le t,\delta_i=1\right)$. Let $\hat{H}_{a}(t)=\left(1-\hat{\eta}\right)^{-1}{\hat{H}}_{1,a}(t)$ which can be expressed as the following product-limit expression: 
\begin{equation*}
\hat{H}_{a}(t)=
\frac{Y_{(t+)}-n\hat{\eta}\hat{G}\left(t\right)}{n\left(1-\hat{\eta}\right)}
=\frac{Y_{(t+)}-n\hat{\eta}\hat{G}\left(t\right)}{R\left(0\right)-n\hat{\eta}\hat{G}\left(0-\right)}
=\prod_{0\le u\le t} \left\{\frac{Y_{(u+)}n\hat{\eta}\hat{G}\left(u\right)}
{Y_{(u)}-n\hat{\eta}\hat{G}\left(u-\right)}
\right\}.
\end{equation*}
Denote $\hat{G}_{a}\left(t\right) = \hat{H}_{a}\left(t\right)/\hat{S}^{PL}_{a}\left(t\right)$. Under the assumption that there are no ties between observed failures and censored data, $\hat{G}_{a}\left(t\right)$ can be expressed as the following product-limit expression: 
\begin{equation*}
{\hat{G}}_{a}\left(t\right)=\prod_{0\le u\le t} \left\{\frac{Y\left(u+\right)-n\hat{\eta}\hat{G}\left(u\right)}{Y\left(u\right)-n\hat{\eta}\hat{G}\left(u-\right)-dN\left(u\right)} \right\} =\prod_{0\le u\le t}\left\{1-\frac{dN_C\left(u\right)+n\hat{\eta}d\hat{G}\left(u\right)}{Y\left(u\right)-n\hat{\eta}\hat{G}\left(u-\right)}\right\},
\end{equation*}
where $N_C\left(t\right)=\sum_{i=1}^{n}I\left(X_i\le t,\delta_i=0\right)$. According to Theorem II.6.1 in \cite{andersen2012statistical}, it is the unique solution to the Volterra integral equation \citep{gill1990survey},
\begin{equation*}
    {\hat{G}}_{a}\left(t\right)=1-\int_{0}^{t}{{\hat{G}}_{a}\left(u-\right)\frac{dN_C\left(u\right)+n\hat{\eta}d\hat{G}\left(u\right)}{Y\left(u\right)-n\hat{\eta}\hat{G}\left(u-\right)}}.
\end{equation*}
By replacing $\hat{G}_{a}(t)$ in the integral equation with $\hat{H}_{a}\left(t\right)/\hat{S}^{PL}_{a}\left(t\right)$, we obtain the following self-consistency equation: 
\begin{equation}
\label{self-consistency2}
    n\left(1-\hat{\eta}\right){\hat{S}}^{PL}_{a}\left(t\right)=\int_{0}^{t}{\hat{\varphi}\left(u\right)\frac{{\hat{S}}^{PL}_{a}\left(t\right)}{{\hat{S}}^{PL}_{a}\left(u-\right)}dN_C\left(u\right)}+n{\hat{H}}_{1,a}\left(t\right),
\end{equation}
which is equivalent to \eqref{self-consistency}.

\section{Proofs of Theorems 2.1, 2.2, and 3.1 under Sufficient Follow-Up}
\label{Proofs}

\subsection{Theorems 2.1: Weak Convergence of $\hat{S}_a(t)$}

To establish the weak convergence of $\sqrt{n}(\hat{S}_{a}(\cdot) - S_{a}(\cdot))$ in the c\`adl\`ag space $D[0,\zeta]$, the following decomposition of $\sqrt{n}(\hat{S}_{a}(\cdot) - S_{a}(\cdot))$ is useful:
\begin{align}
\label{decom_S_hat_a}
     \frac{\sqrt{n}(\hat{S}(\cdot) - S(\cdot)) - \sqrt{n}(\hat{\eta} - \eta)}{1 - \hat{\eta}} + \sqrt{n}(S(\cdot) - \eta)\biggl(\frac{1}{1 - \hat{\eta}} - \frac{1}{1 - \eta}\biggr).
\end{align}
The term 
 $\sqrt{n}(\hat{\eta} - \eta)$ can be treated as a constant function in $D[0, \zeta]$ and is tight in stochastic processes perspective. Together with the tightness of $\sqrt{n}(\hat{S}(\cdot) - S(\cdot))$, the numerator of the first term in \eqref{decom_S_hat_a} is tight. For $0 < \eta < 1$, $\sqrt{n}(\hat{\eta} - \eta)$ converges to a mean-zero normal random variable in distribution \citep{maller1992estimating, maller1996survival}. For $\eta=0$, $\sqrt{n}(\hat{\eta} - \eta) = o_P(1)$. The tightness and pointwise distribution convergence of $\sqrt{n}(\hat{S}(\cdot) - S(\cdot)) - \sqrt{n}(\hat{\eta} - \eta)$ implies the weak convergence in $D[0,\zeta]$. Under the sufficient follow-up assumption, $1 - \eta > 0$, the tightness of the second term of \eqref{decom_S_hat_a} is established since $S(\cdot)$ is assumed continuous. Then, by using Slutsky's theorem and the continuous mapping theorem, we conclude that $\sqrt{n}(\hat{S}_a(\cdot) - S_a(\cdot))$ converges weakly.

\subsection{Theorem 2.2: Efficient Influence Function of $\hat{S}_{a}(t)$} \label{Efficiency}

From the inverse-probability-censoring-weighting  expression $\hat{S}_a^W(t)$ of $\hat{S}_a(t)$, we can write
\begin{equation*}
\sqrt{n} \hat{S}_a(t) =
\begin{aligned}[t]
& \sqrt{n} \left\{ \sum_{i=1}^{n} \frac{I(X_i > t, \delta_i = 1)}{n(1-\hat{\eta}) \hat{G}(X_i)} - \sum_{i=1}^{n} \frac{I(X_i > t, \delta_i = 1)}{n(1-\hat{\eta}) G(X_i)} \right\} \\
&\quad + \sqrt{n} \left\{ \sum_{i=1}^{n} \frac{I(X_i > t, \delta_i = 1)}{n(1-\hat{\eta}) G(X_i)} - \sum_{i=1}^{n} \frac{I(X_i > t, \delta_i = 1)}{n(1-\eta) G(X_i)} \right\} \\
&\quad + \sqrt{n} \sum_{i=1}^{n} \frac{I(X_i > t, \delta_i = 1)}{n(1-\eta) G(X_i)}  \\
&\quad = (a) + (b) + (c).
\end{aligned}
\end{equation*}
By the consistency of $\hat{\eta}$, the first term $(a)$ can be expressed as
\begin{equation*}
(a) = \sqrt{n} \sum_{i=1}^{n} \frac{\delta_iI(X_i > t)}{n(1-\eta) G(X_i)} \left( \frac{G(X_i)}{\hat{G}(X_i)} - 1 \right) + o_p(1).
\end{equation*}
Let $\eta_a = 1 - \eta = \Pr(T \leq t)$ and $\hat{\eta}_a = 1 - \hat{\eta}$. 
The second component  $(b)$, which reflects the influence of $\hat{\eta}$,  can be written as  
\begin{equation*}
(b)  = \frac{S_a(t)}{1-\eta} \sqrt{n} (\hat{\eta} - \eta) + o_p(1)
= \frac{-S_a(t)}{1-\eta} \sqrt{n} (\hat{\eta}_a - \eta_a) + o_p(1).
\end{equation*}
Given that 
\[
\hat{\eta}_a = \sum_{i=1}^{n} \frac{\delta_i}{n\hat{G}(X_i)},
\]
the component $(b)$ can be expressed as 
\begin{equation*}
 -\sqrt{n} S_a (t)   \sum_{i=1}^{n} \frac{\delta_i}{n(1-\eta)G(X_i)}  \frac{G(X_i)}{\hat{G}(X_i)}  
 +\sqrt{n} S_a (t)  +o_p(1).
\end{equation*}
Combining $(a), (b)$ and $(c)$ from the above derivations, we can write 
\[
\sqrt{n} \hat{S}_a(t)
= \sqrt{n}   \sum_{i=1}^{n} \frac{\delta_i \{ I(X_i > t) - S_a(t)\}}
{n(1-\eta)G(X_i)}  \frac{G(X_i)}{\hat{G}(X_i)}  + \sqrt{n} S_a(t)+ o_p(1), 
\]
which can further be simplified as 
\begin{align*}
& \sqrt{n}  \sum_{i=1}^{n} 
\frac{\delta_i \{ I(X_i > t) - S_a(t)\}}{n(1-\eta)G(X_i)} \int_0^{\zeta}
\frac{I(X_i \geq u)}{Y(u)} dM_C(u) \\
& +  \sqrt{n}  \sum_{i=1}^{n} 
\frac{\delta_i \{ I(X_i > t) - S_a(t)\}}{n(1-\eta)G(X_i)} + \sqrt{n} S_a(t) + o_p(1) \\
& = 
n^{-1/2} \left\{ \sum_{i=1}^{n} \int_0^{\zeta}
\frac{ I(X_i \geq u,\delta_i =1) \{ I(X_i > t) - S_a(t)\}}{n(1-\eta)G(X_i)} \frac{n}{Y(u)} \right\}
 \left\{ \sum_{j=1}^n dM_{C,j}(u) \right\}
\\
& + \sqrt{n}  \sum_{i=1}^{n} 
\frac{\delta_i \{ I(X_i > t) - S_a(t)\}}{n(1-\eta)G(X_i)} + \sqrt{n} S_a(t) + o_p(1) \\
& =  n^{-1/2} \sum_{j=1}^{n} \int_{0}^{\zeta} \frac{q_t(u)}{\pi(u)} dM_{C,j}(u) + \sqrt{n}  \sum_{i=1}^{n} 
\frac{\delta_i \{ I(X_i > t) - S_a(t)\}}{n(1-\eta)G(X_i)} + \sqrt{n} S_a(t) +o_p(1),  
\end{align*}
where $M_C(t) = \sum_{j=1}^n M_{C,j}(t)$, 
$
M_{C,j}(t) =  I(X_j \leq t, \delta_j = 0) - \int_0^t I(X_j \geq u) \lambda_c(u)du$,
$\lambda_c(u)$ is the hazard function of $C$ at time $u$,
$q_t(u)$ is the probability limit of 
\[
\sum_{i=1}^{n} \frac{I(X_i\geq u, \delta_i=1) [I(X_i>t) - S_a(t)]}{n(1-\eta)G(X_i)}
\]
and  $\pi(u)$ is the probability limit of $Y(u)/n$. Finally, we have 
\begin{equation} \label{combinedterms}
    \hat{S}_{a}(t) - S_{a}(t) = \frac{1}{n} \sum_{i=1}^{n} \left\{ \frac{\delta_i \left\{I(X_i>t) - S_a(t)\right\}}
{(1-\eta)G(X_i)}  + \int_0^{\zeta} \frac{q_t(u)}{\pi(u)} dM_{C,i}(u) \right\} + o_p\left(\frac{1}{\sqrt{n}}\right).
\end{equation}
The influence function of $\hat{S}_{a}(t)$ can be written as 
\begin{equation} \label{influence_function}
    \psi_t(X, \delta) = \frac{\delta \left\{I(X>t) - S_a(t)\right\}}
{(1-\eta)G(X)}  + \int_0^{\zeta} \frac{q_t(u)}{\pi(u)} dM_{C,1}(u). 
\end{equation}
The asymptotic normality follows directly from the Central Limit Theorem for the sum of identically and independently distributed random variables.

The asymptotic efficiency of $\hat{S}_{a}(t)$ is established in the nonparametric setting \citep[Section VIII.2.4]{andersen2012statistical} through the representation given in \eqref{combinedterms}, which involves the efficient influence function \citep[Theorem 25.20]{vaart_1998}.

  \subsection{Theorem 3.1: Asymptotic Properties of $\hat{\tau}_a(t)$}
We derive the asymptotic properties of $\sqrt n\left({\hat{\tau}}_a\left(t\right)-\tau_a\left(t\right)\right)$. Let 
\begin{equation*}
    \hat{\tau}_{a,1}(t) = \frac{\sum_{i,j}\hat{\psi}_{ij}(t)\hat{w}_{ij}}{n_0 n_1 (1 - \eta_0) (1 - \eta_1)}, 
\end{equation*}

\begin{equation*}
    \hat{\tau}_{a,2}(t) = \frac{\sum_{i,j}\psi_{ij}(t)\hat{w}_{ij}}{n_0 n_1 (1 - \eta_0) (1 - \eta_1)}, 
\end{equation*}
and 
\begin{equation*}
    \hat{\tau}_{a,3}(t) = \frac{\sum_{i,j}\psi_{ij}(t)w_{ij}}{n_0 n_1 (1 - \eta_0) (1 - \eta_1)} .
\end{equation*}
Then we can write 
\begin{align*}
    &\sqrt{n}(\hat{\tau}_a(t) - \tau_a(t)) \\
    =& \sqrt{n}(\hat{\tau}_{a}(t) - \hat{\tau}_{a,1}(t)) + \sqrt{n}(\hat{\tau}_{a,1}(t) - \hat{\tau}_{a,2}(t)) + \sqrt{n}(\hat{\tau}_{a,2}(t) - \hat{\tau}_{a,3}(t)) + \sqrt{n}(\hat{\tau}_{a,3}(t) - \tau_{a}(t)) \\
    =& (A) + (B) + (C) + (D).
\end{align*}

For component (A),
\begin{equation*}
    \sqrt{n}(\hat{\tau}_{a}(t) - \hat{\tau}_{a,1}(t)) = \sqrt{n}\left[1 - \frac{(1 - \hat{\eta}_0) (1 - \hat{\eta_1})}{(1 - \eta_0)(1 - \eta_1)}\right]\frac{\sum_{i,j}\hat{\psi}_{ij}(t) \hat{w}_{ij}}{n_0 n_1 (1 - \hat{\eta}_0) (1 - \hat{\eta}_1)},
\end{equation*}
where
\begin{align*}
    \sqrt{n}\left[1 - \frac{(1 - \hat{\eta}_0) (1 - \hat{\eta_1})}{(1 - \eta_0)(1 - \eta_1)}\right] 
    &= \sqrt{n} \left[1 - \frac{(1 - \hat{\eta}_0) (1 - \eta_1)}{(1 - \eta_0) (1 - \eta_1)} + \frac{(1 - \hat{\eta}_0) (1 - \eta_1)}{(1 - \eta_0) (1 - \eta_1)} - \frac{(1 - \hat{\eta}_0) (1 - \hat{\eta}_1)}{(1 - \eta_0) (1 - \eta_1)}\right] \\
    &= \sqrt{\frac{n}{n_0}}\frac{\sqrt{n_0} (\hat{\eta}_0 - \eta_0)}{1 - \eta_0} + \frac{(1 - \hat{\eta}_0)}{(1 - \eta_0) (1 - \eta_1)}\sqrt{\frac{n}{n_1}}\sqrt{n_1}(\hat{\eta_1} - \eta_1) \\
    &= p_0^{-1/2}(1 - \eta_0)^{-1} \sqrt{n_0}(\hat{\eta}_0 - \eta_0) + p_1^{-1/2}(1 - \eta_1)^{-1} \sqrt{n_1} (\hat{\eta}_1 - \eta_1) + o_P(1).
\end{align*}
Hence, the limiting behavior of (A) depends on $\sqrt{n_\ell}(\hat{\eta}_\ell - \eta_\ell),\ \ell=0,1$, under sufficient follow-up (\cite{maller1996survival}: Theorem 4.3). For $\eta_\ell=0$, $\sqrt{n_\ell}(\hat{\eta}_\ell - \eta_\ell) = o_P(1)$. For $0 < \eta_\ell < 1$, $\sqrt{n_\ell}(\hat{\eta}_\ell - \eta_\ell)$ converges to a Normal random variable in distribution. Accordingly, 
\begin{equation*}
    \sqrt{n}(\hat{\tau}_{a}(t) - \hat{\tau}_{a,1}(t)) = [p_0^{-1/2}(1 - \eta_0)^{-1} \sqrt{n_0}(\hat{\eta}_0 - \eta_0) + p_1^{-1/2}(1 - \eta_1)^{-1} \sqrt{n_1} (\hat{\eta}_1 - \eta_1)] \tau_a(t) + o_P(1).
\end{equation*}

For component (B),
\begin{equation*}
    \sqrt{n}(\hat{\tau}_{a,1}(t) - \hat{\tau}_{a,2}(t)) = \sqrt{n} \sum_{i,j}{\left[1 - \frac{\hat{G}_0(\tilde{X}_{ij}) \hat{G}_1(\tilde{X}_{ij})}{G_0(\tilde{X}_{ij}) G_1(\tilde{X}_{ij})}\right] \frac{\hat{\psi}_{ij}(t) \hat{w}_{ij}}{n_0 n_1 (1 - \eta_0) (1 - \eta_1)}}, 
\end{equation*}
where 
\begin{align*}
    1 - \frac{\hat{G}_0(\tilde{X}_{ij}) \hat{G}_1(\tilde{X}_{ij})}{G_0(\tilde{X}_{ij}) G_1(\tilde{X}_{ij})} &= \frac{G_0(\tilde{X}_{ij}) G_1(\tilde{X}_{ij}) - \hat{G}_0(\tilde{X}_{ij}) \hat{G}_1(\tilde{X}_{ij})}{G_0(\tilde{X}_{ij}) G_1(\tilde{X}_{ij})} \\
    &= \frac{G_0(\tilde{X}_{ij}) G_1(\tilde{X}_{ij}) - \hat{G}_0(\tilde{X}_{ij}) G_1(\tilde{X}_{ij}) + \hat{G}_0(\tilde{X}_{ij}) G_1(\tilde{X}_{ij}) - \hat{G}_0(\tilde{X}_{ij}) \hat{G}_1(\tilde{X}_{ij})}{G_0(\tilde{X}_{ij}) G_1(\tilde{X}_{ij})} \\
    &= \left[1 - \frac{\hat{G}_0(\tilde{X}_{ij})}{G_0(\tilde{X}_{ij})}\right] + \frac{\hat{G}_0(\tilde{X}_{ij})}{G_0(\tilde{X}_{ij})}\left[1 - \frac{\hat{G}_1(\tilde{X}_{ij})}{G_1(\tilde{X}_{ij})}\right] \\
    &= \int_{0}^{\tilde{X}_{ij}}{\frac{\hat{G}_0(u-)}{G_0(u) Y_0(u)} dM_0^C(u)} + \frac{\hat{G}_0(\tilde{X}_{ij})}{G_0(\tilde{X}_{ij})} \int_{0}^{\tilde{X}_{ij}}{\frac{\hat{G}_1(u-)}{G_1(u) Y_1(u)} dM_1^C(u)},
\end{align*}
 $\hat{G}_\ell(t)$ is the Kaplan-Meier estimator of $\mbox{Pr}(C_\ell > t)$, $M_\ell^C(t)$ and $Y_\ell(t)$ represent the group-$\ell$ versions of $M_C(t)$ and $Y(t)$, respectively, for $\ell=0,1$. 
The last equation follows by the martingale integral representation of Kaplan-Meier estimator (\cite{andersen2012statistical}, p.257).  
Then, by the Martingale Central Limit Theorem (Theorem II.5.1 in \cite{andersen2012statistical}),
\begin{align*}
    \sqrt{n}(\hat{\tau}_{a,1}(t) - \hat{\tau}_{a,2}(t)) &= \sqrt{n} \sum_{i,j}{\int_{0}^{\infty}{\frac{\hat{G}_0(u-)}{G_0(u) Y_0(u)} \frac{\hat{\psi}_{ij}(t) \hat{w}_{ij} I(u \le \tilde{X}_{ij})}{n_0 n_1 (1 -\eta_0) (1 - \eta_1)}dM_0^C(u)}} \\
    &\quad + \sqrt{n} \sum_{i,j}{\frac{\hat{G}_0(\tilde{X}_{ij})}{G_0(\tilde{X}_{ij})} \int_{0}^{\infty}{\frac{\hat{G}_1(u-)}{G_1(u) Y_1(u)} \frac{\hat{\psi}_{ij}(t) \hat{w}_{ij} I(u \le \tilde{X}_{ij})}{n_0 n_1 (1 -\eta_0) (1 - \eta_1)}dM_1^C(u)}} \\
    &= p_1^{-1/2}n_0^{-1/2} \int_{0}^{\infty}{\frac{\tau_a(u,t)}{\Pr(X_0 \ge u)} dM_0^C(u)} + p_0^{-1/2} n_1^{-1/2} \int_{0}^{\infty}{\frac{\tau_a(u,t)}{\Pr(X_1 \ge u)} dM_1^C(u)} + o_P(1),
\end{align*}
where 
\begin{equation*}
    \tau_a(u,t) = E\left[sign(X_{1,j} - X_{0,i})I(u \le \tilde{X}_{ij} \le t) \Big| \xi_{0,i} = \xi_{1,j} = 1\right].
\end{equation*}

For component (C), we can write  
\begin{equation*}
    \sqrt{n}(\hat{\tau}_{a,2}(t) - \hat{\tau}_{a,3}(t)) = \sqrt{n} \sum_{i,j}(\hat{w}_{ij} - w_{ij}) \frac{\psi_{ij}(t)}{n_0 n_1 (1 - \eta_0)(1 - \eta_1)}.
\end{equation*}
For the case with $\delta_{0,i} = 0, \delta_{1,j} = 1$, it follows that 
\begin{equation*}
    \hat{w}_{ij} - w_{ij} = \frac{(1 - \hat{\eta}_0)\hat{S}_{a,0}(X_{0,i})}{(1 - \hat{\eta}_0)\hat{S}_{a,0}(X_{0,i})-\hat{\eta}_0} - \frac{(1 - \eta_0)S_{a,0}(X_{0,i})}{(1 - \eta_0)S_{a,0}(X_{0,i}) - \eta_0}.
\end{equation*}
Define $M_\ell(t)$ as the martingale for $N_\ell(t) = \sum_{i=1}^{n_\ell} I(X_{\ell,i} \leq t, \delta_{\ell,i}=1)$ and $\hat{S}_\ell(t)$ as the Kaplan-Meier estimator of $S_\ell(t)$, where $\ell=0,1$. 
After simplifying to a common denominator, the numerator of $\hat{w}_{ij} - w_{ij}$ can be expressed as
\begin{align*}
    &S_{a,0}(X_{0,i})(\hat{\eta}_0 - \eta_0) - (1 - \hat{\eta}_0)\eta_0[\hat{S}_{a,0}(X_{0,i}) - S_{a,0}(X_{0,i})] \\
    =& -\eta_0 S_{a,0}(X_{0,i})\int_{0}^{X_{(n_0)}}\frac{\hat{S}_0(u-)}{S_0(u)}\frac{J_0(u)}{Y_0(u)}dM_0(u) + (1 - \hat{\eta}_0) \eta_0 S_{a,0}(X_{0,i}) \int_{0}^{X_{0,i}}\frac{\hat{S}_0(u-)}{S_0(u)}\frac{J_0(u)}{Y_0(u)}dM_0(u),
\end{align*}
by the property of the Kaplan-Meier estimator, the asymptotic property of $\hat{S}_a(t)$ stated in Section 2.3 and the asymptotic property of $\hat{\eta}_j$. 
In other cases, we can derive similar expressions using this technique. Subsequently, component (C) converges in distribution to a normal random variable by the martingale central limit theorem.

For component (D), its distributional convergence can be achieved using U-statistic theory, as discussed in
 \citet[Chapter 12]{vaart_1998}.

\section{Heuristic Analysis of Estimators Under Insufficient Follow-Up}
\label{Insufficient_Follow_Up}
Under conditions of insufficient follow-up, we employ $\check{S}_a^{LS}(t;b)$, as defined in Section 2.3 of the main text, to estimate $S_a(t)$, where the parameter $\eta$ is estimated using $\check{\eta}_b$, as proposed by Escobar-Bach and Van Keilegom \cite{EscobarBach2019}.  Given that $\check{S}_a^{LS}(t;b)$ is a linear function of the Kaplan-Meier estimator $\hat{S}(t)$ and the modified cure rate estimator $\check{\eta}_b$, it inherits their probabilistic properties when the required conditions are satisfied. It is important to note that although $\check{\eta}_b$ reduces the bias of $\hat{\eta} = \hat{S}(t_{K})$ as an estimator of $\eta$, it does not achieve consistency.

Here we analyze the bias of $\check{S}_a^{LS}(t;b)$ as an estimator of $S_a(t)$. Asymptotically, we can treat \(\hat{S}(t)\) as an unbiased estimator of \(S(t)\). 
Define 
\[ f(\eta; S(t)) = \frac{S(t) - \eta}{1 - \eta}, \]
the first derivative of \( f \) with respect to \(\eta\) is given by:
\[ f'(\eta; S(t)) = \frac{S(t) - 1}{(1 - \eta)^2}. \]
Let \(\theta\) and \(\theta'\) represent the large-sample biases of \(\hat{\eta}\) and \(\check{\eta}_b\), respectively, in estimating \(\eta\). To evaluate the biases of \(\hat{S}_a(t)\) and \(\check{S}_a^{LS}(t;b)\) in estimating \(S_a(t)\), we compute \(f(\eta + \theta)\) for \(\hat{S}_a(t)\) and \(f(\eta + \theta')\) for \(\check{S}_a^{LS}(t;b)\) and obtain:
\begin{align*}
\text{Bias of } \hat{S}_a(t) &\approx \frac{S(t) - 1}{(1 - \eta)^2} \cdot \theta, \\
\text{Bias of } \check{S}_a^{LS}(t;b) &\approx \frac{S(t) - 1}{(1 - \eta)^2} \cdot \theta'.
\end{align*}
Given \( |\theta'| < |\theta| \), \(\check{S}_a^{LS}(t;b)\) is less biased than \(\hat{S}_a(t)\) in estimating \(S_a(t)\).
In Equation (18) of the main text describing $\hat{\tau}_a(t)$, cure rate estimators are utilized not only in the normalization component but also in the weighting factors. This complexity makes it difficult to directly compare $\hat{\tau}_a(t)$ with $\check{\tau}_a(t; b_0, b_1)$. According to our data analysis in Section 5 of the main text, $\hat{\tau}_a(t)$ and $\check{\tau}_a(t; b_{*,0}, b_{*,1})$ exhibit similar shapes.

\section{Simulation Analysis for $\hat{S}_a(t)$ and $\hat{\eta}$}
\label{simulations}
The finite-sample performance of $\hat{S}_a(t)$ and $\hat{\eta}$ are evaluated under two different cure rates, $\eta = 0.2$ and $0.4$. 
There are two regularity conditions for Theorem 1: $\zeta \leq \zeta_C$ and 
 \begin{equation}
 \label{regularity1}
        \int_0^{\zeta}\frac{dF(u)}{G(u-)}<\infty.
  \end{equation}
We investigate the performances of $\hat{S}_a(t)$ under 
the following three settings:
\begin{itemize}
    \item \textbf{Case 1:} $\zeta < \zeta_C$;
    \item \textbf{Case 2:} $\zeta = \zeta_C$ but \eqref{regularity1} does not hold;
    \item \textbf{Case 3:} $\zeta = \zeta_C$ and \eqref{regularity1} holds. 
\end{itemize}
Note that when $\zeta < \zeta_C$, the condition \eqref{regularity1} is satisfied. Therefore, Theorem 1 applies to Cases 1 and 3, while Case 2 describes the situation when the regularity condition is violated. We present the results of Case 3 with $n=200$ in Table 1 of the main text and the results with $n=400$ along with additional information in this supplement. 

Simulation results for Case 1 and Case 2 are presented in Section 1 and Section 2, respectively. We generate $T|\xi = 1$ such that $S_a(t) = \Pr(T>t|T \leq t_b)$, where $T$ follows a Weibull distribution with a shape parameter of $0.75$, a scale parameter of $1.5$, and two different truncation points $t_b$.  In all settings, the censoring variable $C$ is generated from uniform distributions with different lengths of support.  Figure~\ref{Fig_Weibull_3} displays $G(t) = \Pr(C>t)$ and the density function of $T|\xi = 1$ together, allowing for a visual check of the validity of \eqref{regularity1}. The left plot of  Figure~\ref{Fig_Weibull_3} 
corresponds to Case 2, where $C \sim \text{Uniform}[0, 4]$ and $ t_b = 4$.  The right plot  corresponds to Case 1, where $C \sim  \text{Uniform}[0, 5]$ and $ t_b = 4$.   

In each table, we provide the average bias of $\hat{\eta}$ and $\hat{S}_a(t)$ at selected time points $t$, the standard deviation estimates obtained from 2000 bootstrap resamples, and the empirical coverage probabilities for the estimated confidence intervals. These performance metrics are calculated based on 500 simulation runs.

\subsection{Simulation results for Case 1}
Case 1 is shown in the right plot of Figure~\ref{Fig_Weibull_3}, which indicates that equation \eqref{regularity1} holds, and Figure~\ref{Fig_Weibull_1_small}.  
Table~\ref{table_trun_Weibull_1} presents the results based on a sample size of $n = 200$. The average biases are close to zero, and the bootstrap estimates of the standard deviation of $\hat{S}_a(t)$ closely match the empirical estimates. Additionally, the empirical coverage probabilities are approximately at the 95\% nominal level. Notice that the confidence intervals are wider under $\eta = 0.4$ compared to $\eta = 0.2$.

\begin{center}
[Insert Figures~\ref{Fig_Weibull_3}, ~\ref{Fig_Weibull_1_small} and Table~\ref{table_trun_Weibull_1}]
\end{center}

\subsection{Simulation results for Case 2}
 This case is shown in the left plot of Figure~\ref{Fig_Weibull_3}, which 
 indicates that  equation  \eqref{regularity1} is violated, as it is still possible to observe a failure as $t$ approaches $\zeta_C = \zeta$.
The simulation results are presented in Table~\ref{table_trun_Weibull_2}. Although most results still appear to be reasonable, the performance of $\hat{\eta}$ and $\hat{S}_a(t)$ for large $t$ becomes worse compared to their counterparts in Table~\ref{table_trun_Weibull_1}. In addition, the empirical coverage probabilities become less accurate in Case 2. 

\begin{center}
[Insert Table~\ref{table_trun_Weibull_2}]
\end{center}

\subsection{Simulation results under $\zeta = \zeta_C$ and \eqref{regularity1} holds}
We generate $T|\xi = 1$ from a Beta distribution with parameters $\alpha_1 = 1$ and $\alpha_2 = 3$. The censoring variable $C$ follows $\text{Uniform}[0, 1]$. As shown in Figure~\ref{Fig_Beta_dist_1_3}, we observe that the condition specified in \eqref{regularity1} is satisfied. The results based on a sample size of $n=200$ are presented in Table 1 of the main text.   Here in  Table~\ref{table_Beta_2}, we present the results with  $n=400$.  We see that the confidence interval lengths become smaller compared to those in Table 1 of the main text.

\begin{center}
[Insert Figure~\ref{Fig_Beta_dist_1_3} and Table~\ref{table_Beta_2}]
\end{center}

\section{Simulation Analysis for $\hat{\tau}_a(t)$}
We examine the performance of $\hat{\tau}_a(t)$ under different scenarios of $S_{a,\ell}(t)$ and $\eta_\ell$ for $\ell=0,1$.

\subsection{Non-crossing Susceptible Survival Curves}
We present two examples where the susceptible survival functions do not intersect and  $\eta_0=\eta_1=0.2$, as shown in Figures~\ref{Fig_tau_plot_1} and \ref{Fig_tau_plot_2}. In the main text, we present the simulation results for the latter case. This scenario demonstrates a greater disparity between the two susceptible groups, resulting in a higher curve $\tau_a(t)$.
The simulation results for the former case are given in Table~\ref{tau_no_crossing_1}. 
We observe that the average bias of $\hat{\tau}_a(t)$ is almost zero. The bootstrap estimates of the standard deviation of $\hat{\tau}_a(t)$ closely align with the empirical estimates, and the empirical coverage probabilities are close to the 95\% nominal level.  Notice that the lengths of the confidence intervals are wider under $\eta_0=\eta_1 = 0.4$ when compared to those under  $\eta_0=\eta_1 = 0.2$. According to the last column of Table~\ref{tau_no_crossing_1}, $\hat{\tau}_a(1)$ does not yield a significant result for testing $H_0: \tau_a(1) = 0$.

\begin{center}
[Insert Figures~\ref{Fig_tau_plot_1}, \ref{Fig_tau_plot_2} and Table~\ref{tau_no_crossing_1} ]
\end{center}

\subsection{Crossing Susceptible Survival Curves}
Figures~\ref{Fig_tau_plot_3} and ~\ref{Fig_tau_plot_4} depict scenarios in which the susceptible survival functions intersect at a susceptible survival probability approximately around 0.5, with $(\eta_0,\eta_1)= (0.2,0.2)$ and $(0.2,0.4)$, respectively. 
The corresponding curve $\tau_a(t)$ initially decreases before ascending later. However in  Figure~\ref{Fig_tau_plot_4}, this reversal phenomenon is not as apparent for $\tau(t)$ since the higher cure rate of Group 1 masks it.

The turning point of $\tau_a(t)$   represents the time of hazard reversal for the susceptibles in the two groups. Finally, $\tau_a(\zeta)$ is only slightly above zero, indicating that the performances of the two susceptible groups are roughly tied. The simulation results of this case are provided in Table 3 of the main text. 

\begin{center}
[Insert Figures~\ref{Fig_tau_plot_3} and ~\ref{Fig_tau_plot_4} ]
\end{center}

\section{Simulation without Cure in One Group}
We also examine the behavior of $\hat{S}_a(t)$ and $\hat{\tau}_a(t)$ in scenarios where there are no cures ($\eta = 0$) in the one-sample case or no cures in at least one group in the two-sample application. The conclusion is that under the specified regularity conditions, these two estimators remain asymptotically normal.

Table~\ref{no_cure} presents the performance of $\hat{S}_a(t)$ and $\hat{\eta}$ when $\eta =0$. The results suggest that $\hat{S}_a(t)$ continues to perform well, but the coverage probability of the 95\% confidence interval based on $\hat{\eta}$ falls below the expected level. 
Figure~\ref{Fig_tau_n0_cure} displays histograms of $\hat{\eta}$ and $\hat{\tau}_a(t)$ at selected values of $t$. These results are based on 2000 runs and 5000 resampling iterations with $n_0=n_1=200$, assuming $T_1|\xi_1 = 1 \sim$ Beta(1,4) and $T_0|\xi_0 = 1 \sim$ Beta(1,2). We see that the normality approximation is satisfactory for $\hat{\tau}_a(t)$ and $\hat{\eta}_1$, but not for $\hat{\eta}_0$.

\begin{center}
[Insert Figure~\ref{Fig_tau_n0_cure} and Table~\ref{no_cure} ]
\end{center}

\section{Additional Data Analysis}
Figure~\ref{fig_cure_riskset} depicts $1 - \hat{\varphi}(t)$ in (10) of the article, the estimated proportion of cured subjects, for the risk sets of the three groups in \cite{hodi2018nivolumab}. The figure reveals that the proportion of cure is significantly higher in the risk set of the combined group than in the group receiving 'ipilimumab alone.' As a result, the hazard ratio would strongly favor the treatment effect of the combined group, which may mitigate the negative aspect of the combined therapy shown in Figures 2(C) and 2(D) of the main text due to patient heterogeneity. 

\begin{center}
[Insert Figure~\ref{fig_cure_riskset} ]
\end{center}

\clearpage
\renewcommand{\thetable}{S\arabic{table}}

\begin{table}[ht]
\centering
\caption{Finite-sample performance of $\hat{S}_a(t)$ and $\hat{\eta}$ based on $n=200$, when $S_a(t)$ follows a truncated Weibull distribution truncated at $t=4$ and $C \sim U(0,5)$. Upper table: $\eta = 0.2$ and $\text{Pr}(\delta = 0) \approx 0.371$; lower table: $\eta = 0.4$ and $\text{Pr}(\delta = 0) \approx 0.528$. (a) Average bias of $\hat{S}_a(t)$; (b) Estimated standard deviation by bootstrap; (c) Empirical standard deviation; (d) Empirical coverage probability of the 95\% confidence interval based on (b); (e) Average length of the 95\% confidence interval.}
\begin{tabular}{@{}ccccccc@{}}
\toprule
$t$ & 0.233 & 0.393 & 0.597 & 0.857 & 1.193 & 1.641 \\
\midrule
$S_a(t)$ & 0.75 & 0.65 & 0.55 & 0.45 & 0.35 & 0.25 \\
\midrule
(a) & -0.002 & -0.001 & -0.002 & -0.002 & -0.001 & -0.001 \\
(b) & 0.036 & 0.041 & 0.044 & 0.046 & 0.046 & 0.045 \\
(c) & 0.036 & 0.041 & 0.045 & 0.046 & 0.047 & 0.047 \\
(d) & 0.943 & 0.950 & 0.948 & 0.942 & 0.942 & 0.934 \\
(e) & 0.141 & 0.159 & 0.172 & 0.179 & 0.181 & 0.177 \\
\midrule
$S_a(t)$ & 0.75 & 0.65 & 0.55 & 0.45 & 0.35 & 0.25 \\
\midrule
(a) & -0.002 & -0.002 & -0.003 & -0.003 & -0.002 & -0.001 \\
(b) & 0.042 & 0.048 & 0.052 & 0.054 & 0.055 & 0.053 \\
(c) & 0.042 & 0.049 & 0.053 & 0.055 & 0.056 & 0.056 \\
(d) & 0.947 & 0.942 & 0.932 & 0.940 & 0.938 & 0.928 \\
(e) & 0.165 & 0.187 & 0.202 & 0.212 & 0.214 & 0.209 \\
\bottomrule
\end{tabular}
\label{table_trun_Weibull_1}
\end{table}

\begin{table}[ht]
\centering
\caption{Finite-sample performance of $\hat{S}_a(t)$ and $\hat{\eta}$ based on $n=200$, when $S_a(t)$ follows a truncated Weibull distribution truncated at $t=4$ and $C \sim U(0,4)$. Upper table: $\eta = 0.2$ and $\text{Pr}(\delta = 0) \approx 0.413$; lower table: $\eta = 0.4$ and $\text{Pr}(\delta = 0) \approx 0.560$. (a) Average bias of $\hat{S}_a(t)$; (b) Estimated standard deviation by bootstrap; (c) Empirical standard deviation; (d) Empirical coverage probability of the 95\% confidence interval based on (b); (e) Average length of the 95\% confidence interval.}
\begin{tabular}{@{}ccccccc@{}}
\toprule
$t$ & 0.233 & 0.393 & 0.597 & 0.857 & 1.193 & 1.641 \\
\midrule
$S_a(t)$ & 0.75 & 0.65 & 0.55 & 0.45 & 0.35 & 0.25 \\
\midrule
(a) & -0.003 & -0.004 & -0.005 & -0.005 & -0.005 & -0.006 \\
(b) & 0.038 & 0.044 & 0.048 & 0.051 & 0.053 & 0.054 \\
(c) & 0.038 & 0.045 & 0.050 & 0.054 & 0.058 & 0.059 \\
(d) & 0.941 & 0.940 & 0.932 & 0.929 & 0.927 & 0.902 \\
(e) & 0.148 & 0.171 & 0.188 & 0.201 & 0.209 & 0.211 \\
\midrule
$S_a(t)$ & 0.75 & 0.65 & 0.55 & 0.45 & 0.35 & 0.25 \\
\midrule
(a) & -0.003 & -0.004 & -0.006 & -0.006 & -0.005 & -0.006 \\
(b) & 0.044 & 0.052 & 0.057 & 0.061 & 0.063 & 0.064 \\
(c) & 0.044 & 0.053 & 0.061 & 0.065 & 0.069 & 0.071 \\
(d) & 0.946 & 0.939 & 0.927 & 0.923 & 0.914 & 0.906 \\
(e) & 0.174 & 0.202 & 0.223 & 0.239 & 0.249 & 0.250 \\
\bottomrule
\end{tabular}
\label{table_trun_Weibull_2}
\end{table}

\begin{table}[ht]
\centering
\caption{Finite-sample performance of $\hat{S}_a(t)$ and $\hat{\eta}$ based on $n=400$, when $S_a(t)$ follows Beta(1,3) and $C \sim U(0,4)$, with $\eta = 0.2$ resulting in approximately $\text{Pr}(\delta = 0) \approx 0.401$ (upper) and $\eta = 0.4$ resulting in approximately $\text{Pr}(\delta = 0) \approx 0.550$ (lower). (a) Average bias of $\hat{S}_a(t)$; (b) Estimated standard deviation by bootstrap; (c) Empirical standard deviation; (d) Empirical coverage probability of the 95\% confidence interval based on (b); (e) Average length of the 95\% confidence interval.}
\begin{tabular}{@{}cccccccc@{}}
\toprule
$t$ & 0.091 & 0.134 & 0.181 & 0.234 & 0.295 & 0.370 & Cure rate \\
\midrule
$S_a(t)$ & 0.75 & 0.65 & 0.55 & 0.45 & 0.35 & 0.25 & $\eta = 0.2$ \\
\midrule
(a) & -0.001 & -0.001 & -0.001 & -0.001 & -0.001 & -0.002 & 0.001 \\
(b) & 0.026 & 0.029 & 0.032 & 0.033 & 0.034 & 0.033 & 0.030 \\
(c) & 0.026 & 0.030 & 0.032 & 0.033 & 0.034 & 0.034 & 0.029 \\
(d) & 0.941 & 0.941 & 0.939 & 0.949 & 0.949 & 0.932 & 0.955 \\
(e) & 0.101 & 0.115 & 0.124 & 0.130 & 0.132 & 0.128 & 0.117 \\
\midrule
$S_a(t)$ & 0.75 & 0.65 & 0.55 & 0.45 & 0.35 & 0.25 & $\eta = 0.4$ \\
\midrule
(a) & -0.001 & -0.001 & -0.001 & -0.001 & -0.002 & -0.003 & 0.001 \\
(b) & 0.030 & 0.034 & 0.037 & 0.039 & 0.040 & 0.039 & 0.033 \\
(c) & 0.031 & 0.035 & 0.037 & 0.039 & 0.039 & 0.040 & 0.033 \\
(d) & 0.943 & 0.946 & 0.946 & 0.956 & 0.950 & 0.934 & 0.952 \\
(e) & 0.118 & 0.135 & 0.146 & 0.153 & 0.156 & 0.152 & 0.130 \\
\bottomrule
\end{tabular}
\label{table_Beta_2}
\end{table}

\clearpage 

\begin{table}[ht]
\centering
\caption{Finite-sample Performance of $\hat{\tau}_a(t)$, based on $n_0=n_1=200$,  with 
$T_1|\xi_1 = 1 \sim$ Beta(1,3), $T_0|\xi_0 = 1 \sim$ Beta(1,4), $\eta_0 = \eta_1 = 0.2$ (top) and $\eta_0 = \eta_1 = 0.4$ (bottom). (a): Average bias of $\hat{\tau}_a(t)$; (b): Estimated standard deviation by bootstrap; (c): Empirical standard deviation; (d): Empirical coverage probability of the 95\% confidence interval based on (b); (e): Average length of the 95\% confidence interval.}
\label{tau_no_crossing_1}
\begin{tabular}{ccccccccccc}
\hline
& \multicolumn{10}{c}{$\eta_0 = \eta_1 = 0.2, \mbox{Pr}(\delta_0 = 0) = 0.359$, $\mbox{Pr}(\delta_1 = 0) = 0.400$ } \\
\hline
$t$ & 0.1 & 0.2 & 0.3 & 0.4 & 0.5 & 0.6 & 0.7 & 0.8 & 0.9 & 1.0 \\
\hline
$\tau_a(t)$ & 0.075 & 0.113 & 0.131 & 0.139 & 0.142 & 0.143 & 0.143 & 0.143 & 0.143 & 0.143 \\
\hline
(a) & 0.002 & 0.000 & 0.001 & 0.001 & 0.001 & 0.001 & 0.001 & 0.001 & 0.001 & 0.001 \\
(b) & 0.056 & 0.068 & 0.073 & 0.076 & 0.077 & 0.077 & 0.077 & 0.077 & 0.077 & 0.077 \\
(c) & 0.055 & 0.068 & 0.072 & 0.074 & 0.075 & 0.076 & 0.076 & 0.076 & 0.076 & 0.076 \\
(d) & 0.950 & 0.946 & 0.946 & 0.954 & 0.950 & 0.954 & 0.954 & 0.954 & 0.954 & 0.954 \\
(e) & 0.219 & 0.266 & 0.287 & 0.296 & 0.300 & 0.302 & 0.303 & 0.303 & 0.303 & 0.303 \\
\hline
& \multicolumn{10}{c}{$\eta_0 = \eta_1 = 0.4,\mbox{Pr}(\delta_0 = 0) = 0.518$, $\mbox{Pr}(\delta_1 = 0) = 0.552$ } \\
\hline
$t$ & 0.1 & 0.2 & 0.3 & 0.4 & 0.5 & 0.6 & 0.7 & 0.8 & 0.9 & 1.0 \\
\hline
$\tau_a(t)$ & 0.075 & 0.113 & 0.131 & 0.139 & 0.142 & 0.143 & 0.143 & 0.143 & 0.143 & 0.143 \\
\hline
(a) & 0.000 & 0.001 & 0.003 & 0.004 & 0.004 & 0.004 & 0.004 & 0.004 & 0.004 & 0.004 \\
(b) & 0.066 & 0.080 & 0.087 & 0.090 & 0.091 & 0.092 & 0.092 & 0.092 & 0.092 & 0.092 \\
(c) & 0.065 & 0.079 & 0.085 & 0.088 & 0.089 & 0.090 & 0.090 & 0.090 & 0.090 & 0.090 \\
(d) & 0.950 & 0.954 & 0.952 & 0.950 & 0.948 & 0.948 & 0.948 & 0.948 & 0.948 & 0.948 \\
(e) & 0.257 & 0.315 & 0.341 & 0.352 & 0.357 & 0.359 & 0.359 & 0.359 & 0.359 & 0.359 \\
\hline
\end{tabular}
\end{table}

\begin{table}[ht]
\centering
\caption{Finite-sample Performance of $\hat{S}_a(t)$, based on $n=200$,  with 
$T_1|\xi_1 = 1 \sim$ Beta(1,4), $\eta=0$  (a): Average bias of $\hat{\tau}_a(t)$; (b): Estimated standard deviation by bootstrap; (c): Empirical standard deviation; (d): Empirical coverage probability of the 95\% confidence interval based on (b); (e): Average length of the 95\% confidence interval.}
\label{no_cure}
\begin{tabular}{cccccccc}
\toprule
$t$ & 0.069 & 0.102 & 0.139 & 0.181 & 0.231 & 0.293 & Cure rate \\
\midrule
$S_a(t)$ & 0.75 & 0.65 & 0.55 & 0.45 & 0.35 & 0.25 & 0 \\
\hline
(a) & -0.002 & -0.002 & -0.002 & -0.002 & -0.001 & -0.002 & 0.003 \\
(b) & 0.031 & 0.035 & 0.037 & 0.038 & 0.037 & 0.035 & 0.006 \\
(c) & 0.032 & 0.035 & 0.037 & 0.039 & 0.037 & 0.035 & 0.007 \\
(d) & 0.943 & 0.955 & 0.942 & 0.940 & 0.949 & 0.950 & 0.846 \\
(e) & 0.123 & 0.137 & 0.145 & 0.147 & 0.144 & 0.136 & 0.024 \\
\bottomrule
\end{tabular}
\end{table}

\clearpage

\renewcommand{\thefigure}{S\arabic{figure}}

\begin{figure}[ht]
\centering
\includegraphics[width=342pt]{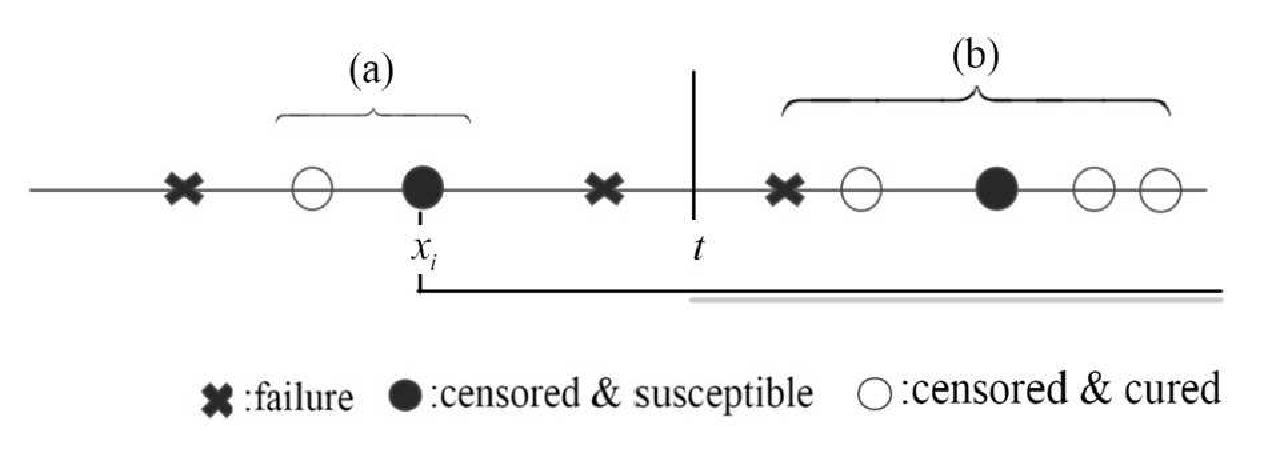}
\caption{Illustration of the proposed self-consistency algorithm. (a): each censored observation weighted by $\hat{\varphi}(X_i^+)\frac{\widetilde{S}_{a}(t)}{\widetilde{S}_{a}(X_i)}$; (b): total number removed: $n \hat{\eta} \hat{G}(t)$.}
\label{fig_self_consistency}
\end{figure}

\begin{figure}[ht]
\centering
\includegraphics{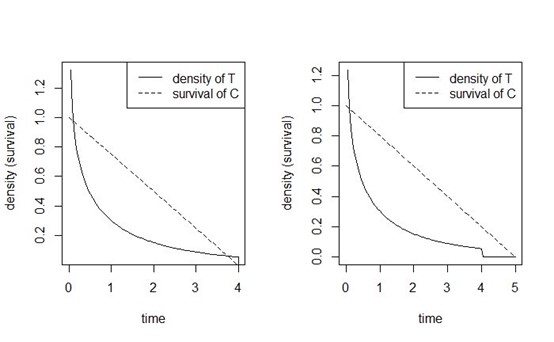}
\caption{Left: Density functions of $T|\xi = 1$ truncated Weibull at $t=4$ and $C$ Uniform in $[0,4]$; Right: Density functions of $T|\xi = 1$ truncated Weibull at $t=4$ and $C$ Uniform in $[0,5]$.}
\label{Fig_Weibull_3}
\end{figure}

\begin{figure}[ht]
\centering
\includegraphics[width=0.8\textwidth]{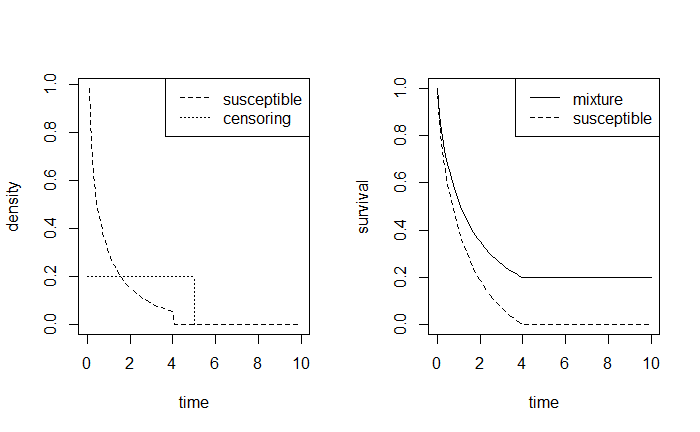}
\caption{Left: Density functions of $T|\xi = 1$ and $C$ truncated Weibull at $t=4$ and Uniform in $[0,5]$; Right: Solid curve represents $S(t)$ and dashed curve represents $S_a(t)$ for $\eta = 0.2$.}
\label{Fig_Weibull_1_small}
\end{figure}


\begin{figure}[ht]
\centering
\includegraphics[width=0.8\textwidth]{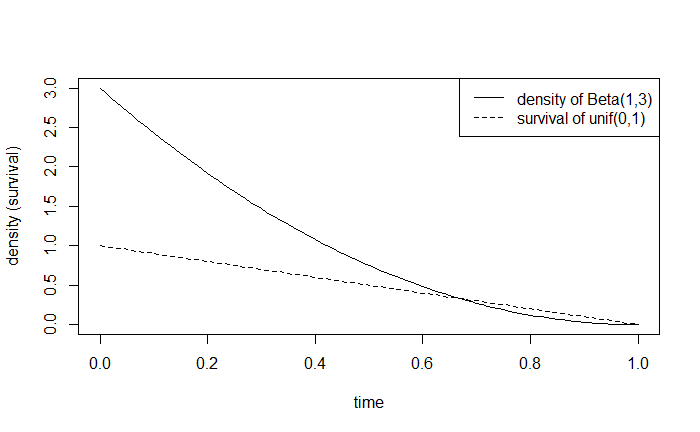}
\caption{Left: Density function of $T|\xi = 1 \sim$ Beta(1,3) and survival function of $C \sim$  Uniform$[0,1]$.}
\label{Fig_Beta_dist_1_3}
\end{figure}

\begin{figure}[ht]
\centering
\includegraphics[width=0.9\textwidth]{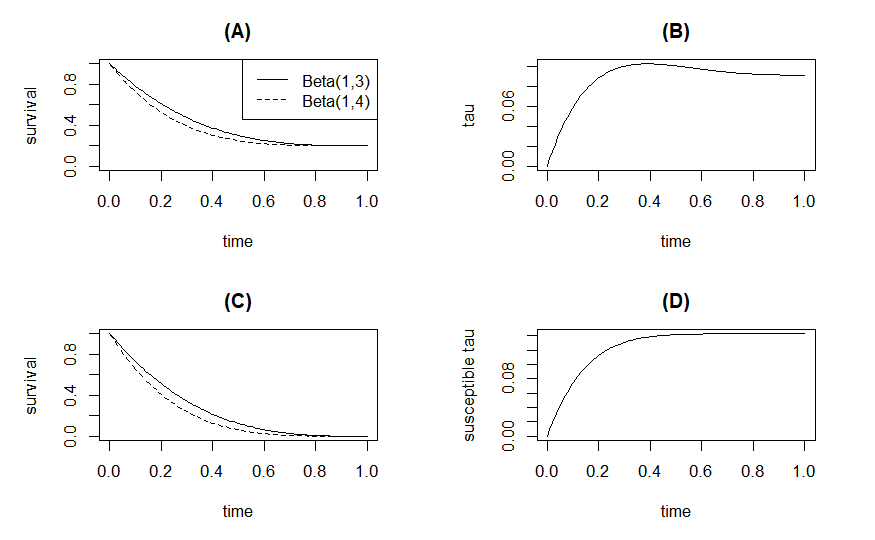}
\caption{(A) Survival functions of $T_\ell$ for $\ell=0,1$ with cure rate 0.2;  (B) $\tau(t)$; (C) Survival functions of $T_1|\xi_1 = 1$ $\sim$ Beta(1,3) (solid) and $T_0|\xi_0 = 1$ $\sim$ Beta(1,4) (dash); (D) $\tau_a(t)$.}
\label{Fig_tau_plot_1}
\end{figure}

\begin{figure}[ht]
\centering
\includegraphics[width=0.9\textwidth]{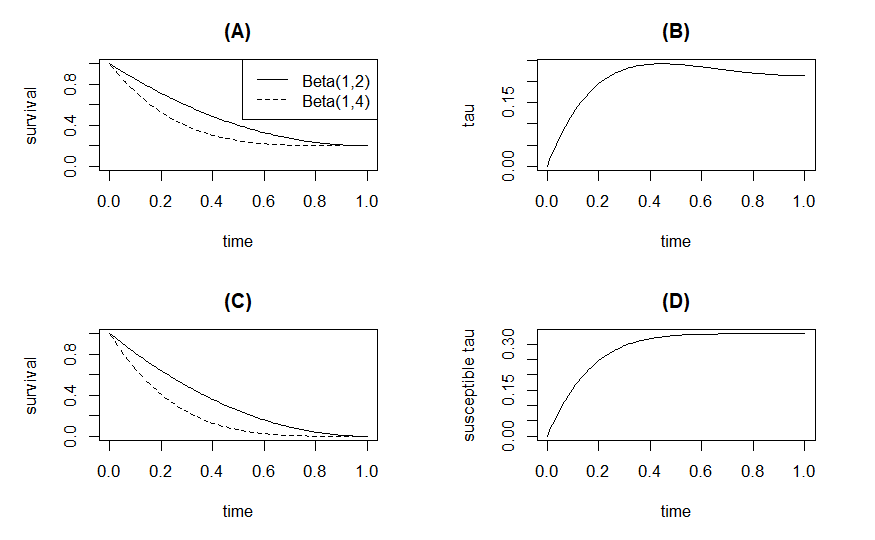}
\caption{(A) Survival functions of $T_\ell$ for $\ell=0,1$ with cure rate 0.2;  (B) $\tau(t)$; (C) Survival functions of $T_1|\xi_1 = 1$ $\sim$ Beta(1,2) (solid) and $T_0|\xi_0 = 1$ $\sim$ Beta(1,4) (dash); (D) $\tau_a(t)$.}
\label{Fig_tau_plot_2}
\end{figure}

\begin{figure}[ht]
\centering
\includegraphics[width=0.9\textwidth]{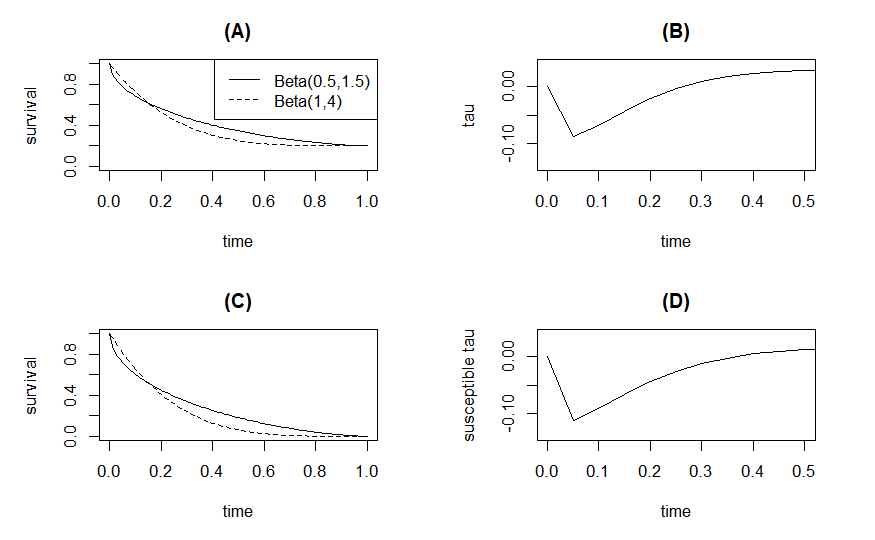}
\caption{(A) Survival functions of $T_\ell$ for $\ell=0,1$ with cure rate 0.2;  (B) $\tau(t)$; (C) Survival functions of $T_1|\xi_1 = 1$ $\sim$ Beta(0.5,1.5) (solid) and $T_0|\xi_0 = 1$ $\sim$ Beta(1,4) (dash); (D) $\tau_a(t)$.}
\label{Fig_tau_plot_3}
\end{figure}

\begin{figure}[ht]
\centering
\includegraphics[width=0.9\textwidth]{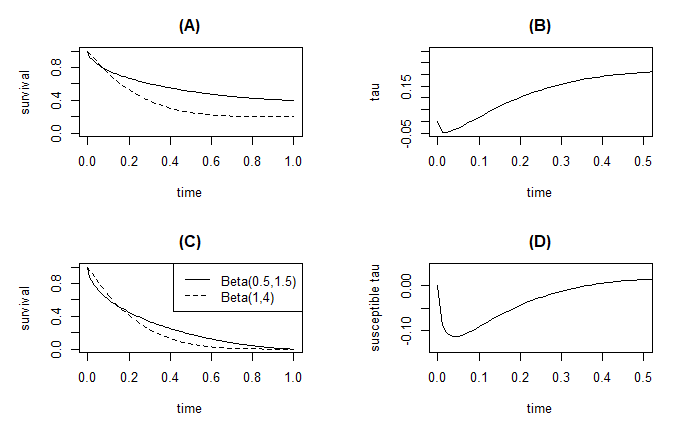}
\caption{(A) Survival functions of $T_0$ and $T_1$ with  $\eta_0 = 0.2, \eta_1 = 0.4$;  (B) $\tau(t)$; (C) Survival functions of $T_1|\xi_1 = 1$ $\sim$ Beta(0.5,1.5) (solid) and $T_0|\xi_0 = 1$ $\sim$ Beta(1,4) (dash); (D) $\tau_a(t)$.}
\label{Fig_tau_plot_4}
\end{figure}

\begin{figure}[ht]
\centering
\includegraphics[width=0.9\textwidth]{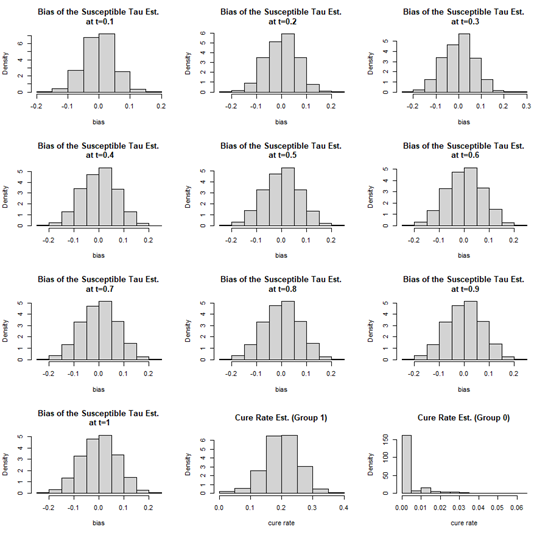}
\caption{Histograms of $\hat{\tau}_a(t)$ and $\hat{\eta}$ with $n_0=n_1=200$, based on 2000 runs and 5000 resampling iterations, where $T_0|\xi_0 = 1 \sim$ Beta(1,4), $T_1|\xi_1 = 1 \sim$ Beta(1,2), $\eta_1 = 0.2$, and $\eta_0 = 0$. Left to right, top to bottom: The first 10 figures represent histograms of the bias of $\hat{S}_a(t)$ at 10 selected values of $t$, while the last two histograms depict $\hat{\eta}_1$ and $\hat{\eta}_0$.}
\label{Fig_tau_n0_cure}
\end{figure}

\begin{figure}[ht]
\centering
\includegraphics[width=0.9\textwidth]{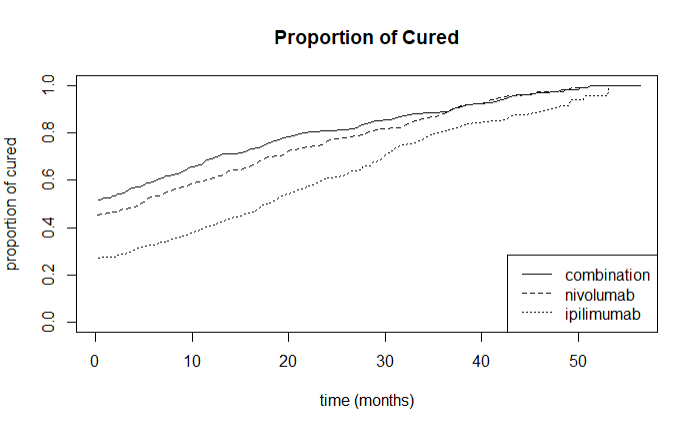}
 \caption{Proportions of cured subjects in the risk sets of "nivolumab plus ipilimumab", "nivolumab alone", and "ipilimumab alone" based on digitized data of CheckMate 067.}
    \label{fig_cure_riskset}
\end{figure}

\clearpage

\bibliography{cure_refs}